\documentclass[english,aps,pra,two column,groupedaddress]{revtex4}
\usepackage[T1]{fontenc}
\usepackage[latin9]{inputenc}
\usepackage{graphicx,array,tabularx}
\usepackage{amstext}
\usepackage{hyperref}
\usepackage{bm,amssymb,amsmath,verbatim,cleveref}



\usepackage{babel}
\usepackage{color}

\begin{document}
\title{Simultaneous model selection and parameter estimation: A superconducting qubit coupled to a bath of incoherent two-level systems}

\author{Markku P.V. Stenberg}
\email[]{markku.stenberg@iki.fi}
\author{Frank K. Wilhelm}
\affiliation{Theoretical Physics, Saarland University, 66123 Saarbr{\"u}cken, Germany}

\begin{abstract}
In characterization of quantum systems, adapting measurement settings based on data while it is collected can generally outperform in efficiency conventional measurements that are carried out independently of data. The existing methods for choosing measurement settings adaptively assume that the model, or the number of unknown parameters, is known. We introduce simultaneous adaptive model selection and parameter estimation. We apply our technique for characterization of a superconducting qubit and a bath of incoherent two-level systems, a leading decoherence mechanism in the state-of-the-art superconducting qubits.
\end{abstract}
\maketitle
\section{Introduction}
Making accurate predictions about nature requires a good model and precise knowledge about its parameters. Model selection \cite{burnhambook,akaike74,schwarz78} and parameter estimation are often considered as different goals but Bayesian inference \cite{sivia06} also provides a unified framework where these two tasks become parts of the same question. 
In this paper, we introduce simultaneous adaptive model selection and parameter estimation. As we explain below, this may be useful, e.g., for characterization of a qubit coupled to a bath of spurious two-level systems, 
a leading decoherence mechanism in the state-of-the-art superconducting qubits.

In solid state devices one can {\it a priori} envision many different sources of decoherence and it is important to find systems where quantum information can be protected from decoherence. 
For a superconducting qubit, a popular early design was based on a Cooper-pair box \cite{nakamura99}, but it was very susceptible to charge noise, a decoherence mechanism that turned out to be extremely difficult to
 eliminate. Reengineering the design in a modified Cooper-pair box, a transmon qubit \cite{koch07}, it was, nevertheless, possible to circumvent the problem and considerably enhance the coherence times \cite{devoret13}.

In modern transmon designs, experiments have provided considerably specific knowledge on the nature of decoherence. A body of experimental evidence suggests that a leading decoherence mechanism is due to a sparse bath of spurious incoherent two-level systems (TLSs) coupled to a qubit. This evidence consists, e.g., on studies of power \cite{schickfus77,martinis05,barends10} and frequency  \cite{barends13} dependence of the dielectric loss as well as on the geometrical location of the TLSs  \cite{gao08,wang09,barends10,wenner11,barends13}. Whereas the qubits with a larger junction area, e.g., phase qubits, often couple to TLSs where an excitation may coherently oscillate between the qubit and a TLS \cite{cooper04,shalibo10}, such coherent TLSs are rare in transmons \cite{barends13} where the junctions are smaller. Despite of these advances, the exact microscopic origin of TLSs is still uncertain, and it is important to gather information on different aspects \cite{palomaki10,lisenfeld10,grabovskij12,sarabi16} of individual TLSs. It is therefore useful to be able to efficiently characterize them. Knowing the TLS frequencies precisely also helps to avoid them in controlling the qubit frequency.

Characterization of quantum systems often assumes that the form of the model is known and only its parameters are uncertain. However, there may also exist significant uncertainty even on the form of the model, i.e., on the number of parameters to be estimated. This is the case, for instance, when a qubit is coupled to TLSs whose number is initially unknown. Within Bayesian inference, model selection and parameter estimation may be carried out simultaneously.  This fact is built in, e.g., in the recently introduced technique of model averaging \cite{ferrie14b}. In \cite{ferrie14b}, random, independently on data chosen, measurements were used for state estimation. While data-independent random controls are more easily realized experimentally and form, e.g., the basis of techniques such as randomized benchmarking \cite{emerson05,knill08,magesan11,magesan12}, it has been shown theoretically \cite{berry09,sergeevich11,huszar12,granade12,ferrie13,ferrie14a,wiebe14a,wiebe14b,stenberg14,granade15,stenberg15,stenberg16} and experimentally  \cite{hannemann02,higgins07,xiang11,yonezawa12,kravtsov13,struchalin16} that in many situations adapting measurement settings during data collection can significantly speed up characterization.

For adaptive measurements, the main task is to assign a set of rules, also called a policy, according to which the measurement settings can be chosen efficiently based on the data obtained. For concreteness, we present such a policy for characterization of a qubit and an initially unknown number of TLSs coupled to it. In the Bayesian framework, information on the previous measurements is, rather than revisiting all the previous data, encoded in the current Bayesian posterior probability distribution which can then for adaptive measurements be used for choosing the measurement settings efficiently. The sequential Monte Carlo (SMC) method \cite{west93,gordon93,liu01} has recently gained  popularity in the quantum context \cite{huszar12,granade12,kravtsov13,ferrie13,wiebe14a,wiebe14b,stenberg14,ferrie14a,stenberg15,stenberg16,struchalin16} due to its computational efficiency. But as we explain below, the existing formulation of SMC does not work when the model is initially unknown. We generalize SMC to account simultaneously both for model selection and parameter estimation. Our approach is not limited to a particular physical system but may be used for a larger class of problems.

\begin{figure}
\includegraphics[width=0.4\textwidth]{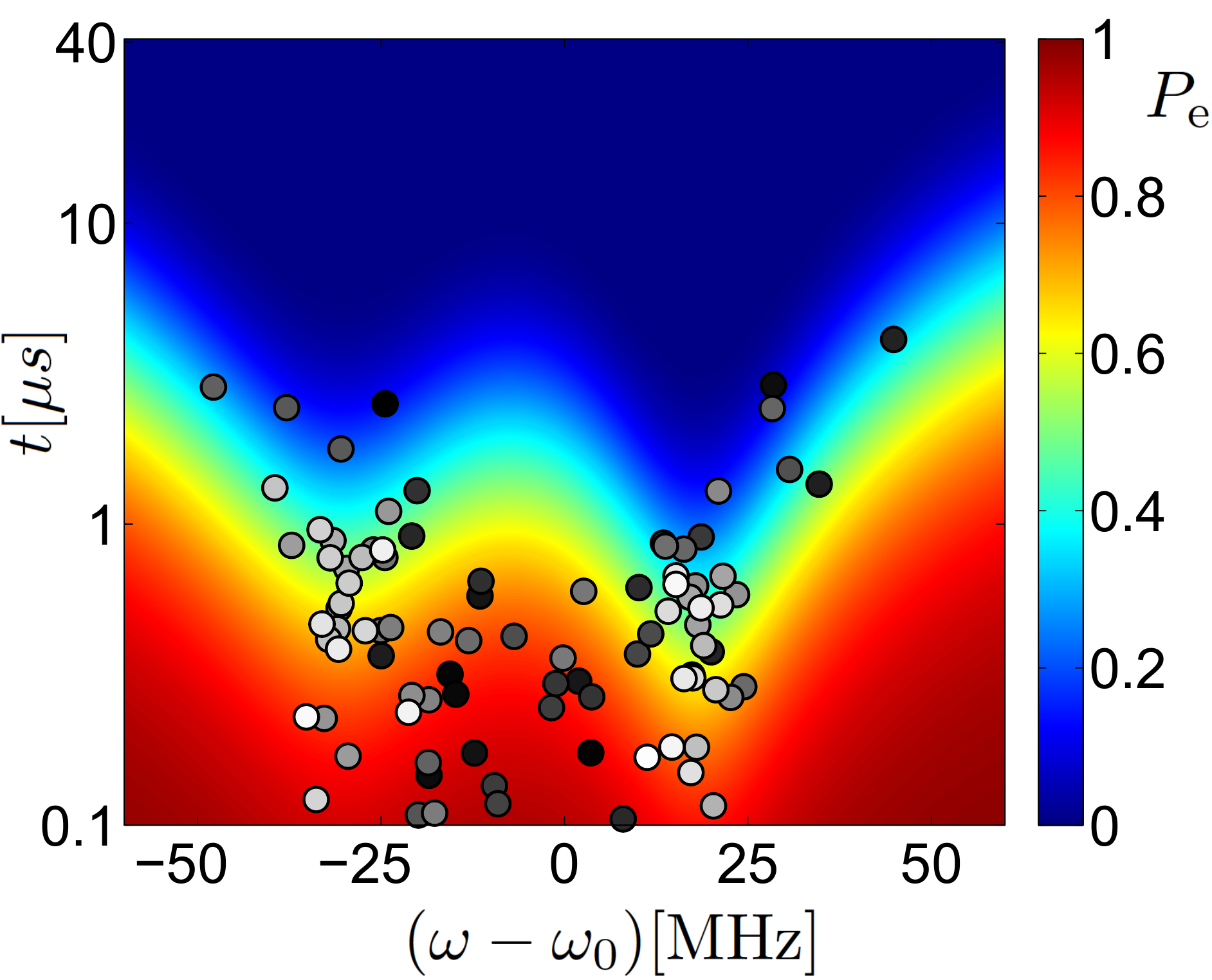}
\caption{The background of the figure exhibits the swap spectrum of two incoherent TLSs coupled to a qubit in a numerically simulated experiment with the model parameters randomly chosen within the region described in Sec.~\ref{sec:iprior}. The color scale for the swap spectrum represents the qubit excited state probability. The filled circles illustrate a simulated sequence of adaptive measurement settings (cf. Sec.~\ref{sec:adpol}) in the frequency--waiting time plane. The order of the measurement settings is denoted by their color (from black to white). For clarity, only every tenth measurement setting is shown. The frequencies are measured with respect to a reference frequency $\omega_0$, cf. Sec.~\ref{sec:iprior}. Note the logarithmic scale on the time axis.}
\label{fig:swap}
\end{figure}

\section{Modeling incoherent two-level systems coupled to a qubit}
\label{sec:model}
A qubit coupled to a bath of TLSs may be described by a master equation in the Lindblad form \cite{lindblad76}
\begin{align}
&\frac{d\hat{\rho}}{dt}=-\frac{i}{\hbar}[\hat{H}_{\rm JC},\hat{\rho}]+\sum_{j=1}^{n_{\rm d}}\sum_{k=1}^{4}\mathcal{D}[\hat{C}_{k}^{(j)}]\hat{\rho},
\label{eq:lindblad}
\end{align}
where $\hat{\rho}$ is the density matrix of the qubit and $n_{\rm d}$ is the number of TLSs coupling to the qubit. When $n_{\rm d}$ is initially unknown, we follow the terminology in the field of model selection and refer to the model as uncertain, even though the form of Eq.~(\ref{eq:lindblad}) is still assumed to be known. The coherent contribution to the time evolution is given by the Hamiltonian \cite{blais04}
\begin{align}
\hat{H}_{{\rm JC}}=&\hbar\omega_{\rm q}\left(\hat{a}^{\dagger}\hat{a}+\frac{1}{2}\right)\nonumber\\
&+\sum_{j=1}^{n_{\rm d}}\left[\frac{\hbar\omega_{\rm d}^{(j)}}{2}\hat{\sigma}_{z}+\hbar g_{j}\left(\hat{\sigma}_{+}^{(j)}\hat{a}+\hat{\sigma}_{-}^{(j)}\hat{a}^{\dagger}\right)\right],
\label{eq:JC}
\end{align}
where $g_j\ll\omega_{\rm d}^{(j)}$. Here, $\omega_{\rm q}$ is the frequency difference of the ground and the first excited state in the qubit (operators $\hat{a}^{(\dag)}$) that is assumed to be controllable in the experimental setup and $\omega_{\rm d}^{(j)}$ is the frequency difference between the two energy levels of the $j$th TLS ($\hat{\sigma}$ operators).
The coupling strength between the qubit and the $j$th TLS is denoted by $g_j$. When the TLSs interact incoherently with the qubit, Markovian decoherence can be described by the Lindblad operator 
described in Appendix~\ref{app:lindblad}.

A qubit coupled to incoherent TLSs undergoes frequency-dependent energy relaxation, so that its excited state probability decreases exponentially \cite{barends13}
\begin{equation}
P_{\rm e}=e^{-\frac{t}{T_1}}.
\label{eq:groundstate_oc}
\end{equation}
The directly observable energy relaxation time $T_1$ of the qubit can be modeled through the frequency-independent energy relaxation time $T_{1,{\rm q}}$ and the decay rates 
$\Gamma_{1,{\rm d}}^{(j)}$ due to different TLSs. For incoherent TLSs, their total contribution is obtained by adding the individual decay rates, and for $\frac{1}{T_{1,Q}}<g_j<
\frac{1}{T_{2,{\rm d}}^{(j)}}$ one obtains \cite{barends13}
\begin{equation}
\frac{1}{T_{1}}=\frac{1}{T_{1,{\rm q}}}+\sum_{j=1}^{n_{\rm d}}\Gamma_{1,{\rm d}}^{(j)},\quad \Gamma_{1,{\rm d}}^{(j)}=\frac{2g_j^2}{1/T_{2,{\rm d}}^{(j)}+T_{2,\rm d}^{(j)}\Delta_j^2}.
\label{eq:incohsum}
\end{equation}
Here, $T_{2,\rm d}^{(j)}$ is the coherence time of the $j$th defect and $\Delta_j=\omega_{\rm q}-\omega_{\rm d}^{(j)}$ is the frequency detuning.

\section{Swap spectroscopy}
The understanding of time resolved spectroscopy, referred to as swap spectroscopy \cite{cooper04,neeley08,shalibo10,mariantoni11a,mariantoni11b,barends13}, may be facilitated by a graphical illustration. Figure~\ref{fig:swap} exhibits a simulated swap spectrum of two incoherent TLSs. One starts by preparing the qubit in the excited state such that the TLSs are in their ground state. This can be done by setting $\omega_{\rm q}$ far from any $\omega_{\rm d}^{(j)}$ and then exciting the qubit with a microwave pulse. Once $\omega_{\rm q}$ has been fixed to a chosen value (horizontal axis in Fig.~\ref{fig:swap}), the system is allowed to evolve a time $t$ (vertical axis in Fig.~\ref{fig:swap}) after which the qubit is measured in the $\hat{\sigma}_{z}$ basis. The system is then reset to its ground state before the next measurement. The measurement is repeated at the same setting $(\omega_{\rm q},t)$ many, usually thousands of times, to approximate the qubit excited state probability (color scale of Fig.~\ref{fig:swap}).  

For coherent TLSs, chevron patterns in the swap spectrum can be used to identify the TLS frequency and the strength of its coupling with the qubit \cite{cooper04}. The conventional 
technique of doing this (for a detailed explanation, see, e.g. \cite{stenberg14}) is based on first finding the TLS frequency $\omega_{\rm d}$ and then deducing the coupling strength
that is proportional to the frequency of coherent oscillations that the excitation undergoes between the qubit and the TLS. This conventional method does not require Bayesian inference but it is not equally efficient or precise than more sophisticated Bayesian schemes \cite{stenberg14,stenberg16}. Furthermore, reviewing the former technique shows that it does not work for incoherent TLSs because an excitation can not coherently oscillate between the qubit and the TLS. Instead, a more general approach based on Bayesian inference can still be applied. 
\section{Adaptive Bayesian inference scheme}
\begin{figure}[h!]
\includegraphics[width=0.4\textwidth]{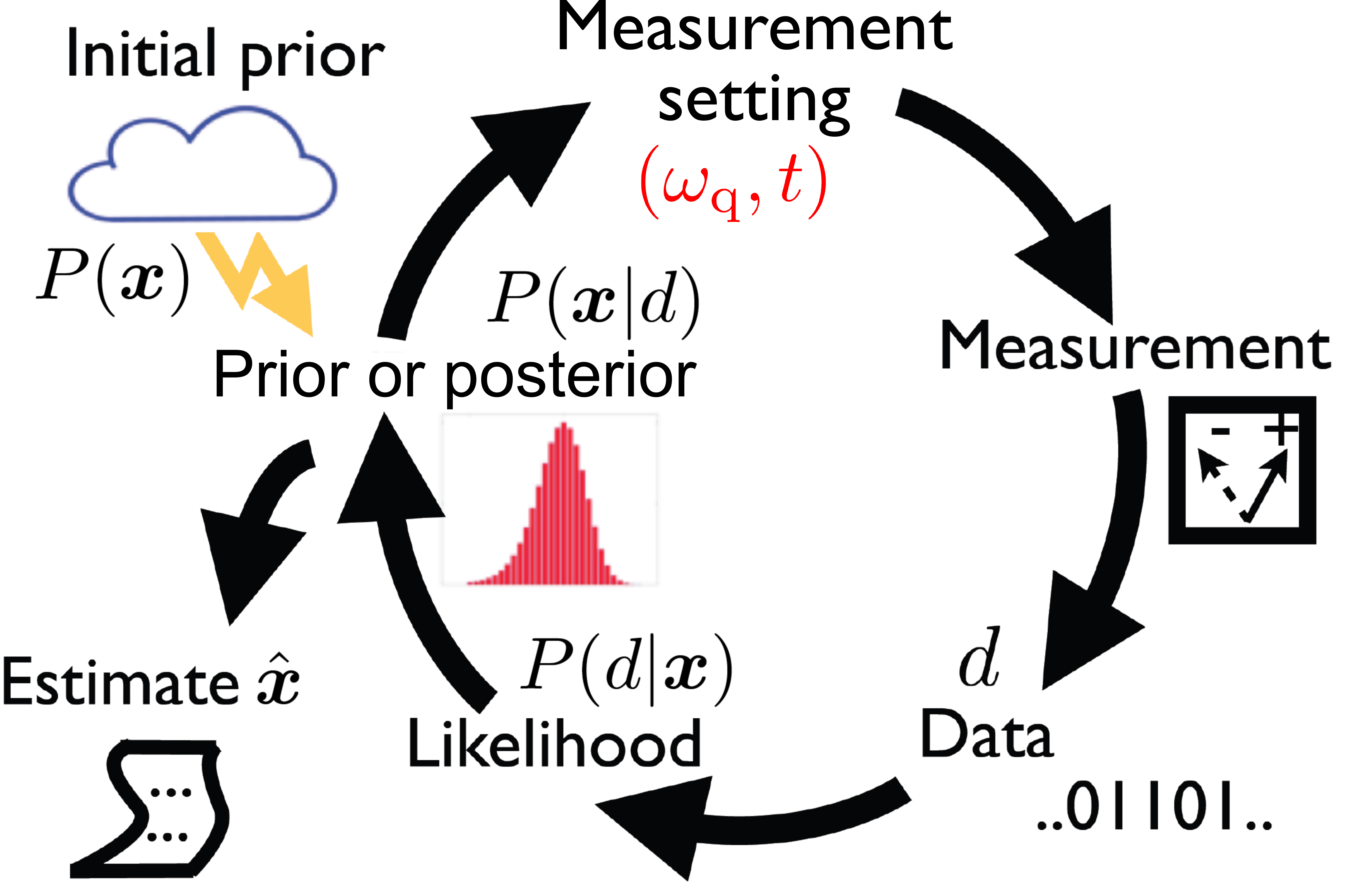}
\caption{Schematics of an adaptive Bayesian inference scheme with the notation used in the main text.}
\label{fig:bayesianinf}
\end{figure}

An adaptive Bayesian inference scheme is illustrated in Fig.~\ref{fig:bayesianinf}. The starting point is the initial prior probability distribution $P({\boldsymbol x}|d_0)$, that describes the experimenters {\it a priori} conception or subjective belief about the values of the unknown parameters encoded in vector ${\boldsymbol x}$ and their uncertainties. In principle, assigning the initial prior does not require any data on the current sample, and one may denote $d_0=\emptyset$. When new data $d_{n+1}$ in the $(n+1)$th measurement setting is obtained, the updated probability 
distribution describing the unknown parameters may be obtained through Bayes' theorem \cite{sivia06}
\begin{equation}
P({\boldsymbol x}|D_{n+1})=\frac{P(d_{n+1}|{\boldsymbol x})P({\boldsymbol x}|D_n)}{\int P(d_{n+1}|{\boldsymbol x}') P({\boldsymbol x}'|D_n) d {\boldsymbol x}'}.
\label{eq:iter_bayes}
\end{equation}
We assume the measurement is repeated at a single setting $M_{\rm r}$ times, so that each $d_n$ denotes the outcome of $M_{\rm r}$ measurement shots.
By a measurement shot we mean a single projective measurement with a binary outcome. The ordered set of measurement outcomes in $n-1$ settings is denoted by  $D_n=(d_0,\ldots,d_n)$. The function $P(d_{n+1}|{\boldsymbol x})$, referred to as likelihood, describes the probability for data $d_{n+1}$ prior to the experiment for different hypothetical parameter values ${\boldsymbol x}$. The likelihood of data $d={\rm e}$ (qubit excited state) is one of the key formulas for the adaptive scheme, and in the setup of this paper it is given by Eqs.~(\ref{eq:groundstate_oc}) and (\ref{eq:incohsum}).

The role of the denominator in Eq.~(\ref{eq:iter_bayes}) is to ensure the normalization dictated by the conservation of probability. The purpose of applying Bayes' theorem is to obtain the quantity $P({\boldsymbol x}|D_{n+1})$, referred to as the posterior, that describes the probability density for different ${\boldsymbol x}$ given data $D_{n+1}$. As illustrated in Fig.~\ref{fig:bayesianinf}, the posterior may be assigned as the prior before the next measurement which makes iterative application of (\ref{eq:iter_bayes}) possible. The estimate is defined by the mean of the posterior $\hat{{\boldsymbol x}}=\int {\boldsymbol x}P({\boldsymbol x}|D)d{\boldsymbol x}$. Here, the hat denotes an estimate rather than an operator, and we have omitted the subscript in $D$ for simplicity. Once a sufficient amount of data has been collected, the estimation scheme is terminated and the final estimate is extracted.

\section{Adaptive detection and characterization of the defects}
\label{sec:adpol}
The Bayesian inference scheme reviewed in the previous section does not, {\it per se}, dictate how the measurement settings should be chosen. By terminology adopted from machine learning, the rules according to which such a choice is made are referred to as a policy. Note that a policy does not necessarily determine the next measurement
setting deterministically but may only give a probability distribution from which the measurement setting is randomly picked. We now present a policy that performs both
model selection and parameter estimation simultaneously.

As explained in Sec.~\ref{sec:model}, we allow the number of spurious TLSs to be initially uncertain. Since characterizing different TLSs generally requires measurements with $\omega_{\rm q}$ set to different frequencies, this raises, e.g., a question, how should the measurements be allocated between different potential TLSs. For incoherent TLSs, we also need a different way of choosing the waiting times $t$ than those discussed, e.g., in \cite{stenberg14,stenberg16} where it was assumed that an excitation may undergo at least some coherent oscillations between the qubit and the TLS. To quantify the difference between coherent and incoherent TLSs, we note that it can be characterized by the ratio $\frac{g_{j}}{\Gamma_{1,{\rm d}}^{(j)}}$. Incoherent TLSs couple incoherently to the qubit and one has $\frac{g_{j}}{\Gamma_{1,{\rm d}}^{(j)}}\lesssim\frac{1}{4}$, implying the absence of coherent oscillations between the qubit and the TLS \cite{barends13}. 

In the absence of an explicit formula for choosing $({\omega_{\rm q}},t)$, an experimenter needs to resort to {\it ad hoc} decisions of his own. To choose $({\omega_{\rm q}},t)$ systematically, we propose the following policy
\begin{align}
& P_{\omega_{\rm q}}=\frac{P_{1,{\rm p}}P_{\omega_{\rm d}^{(1)}}(\omega_{\rm q}|D)+P_{2,{\rm p}}P_{\omega_{\rm d}^{(2)}}(\omega_{\rm q}|D)}{P_{1,{\rm p}}+P_{2,{\rm p}}},\nonumber\\
&t=r\hat{T_1},\quad r\in (0,1].
\label{eq:P_omega_q}
\end{align}
Here, $\omega_{\rm q}$ is chosen randomly following the probability distribution $P_{\omega_{\rm q}}$. The latter is a linear combination of posteriors $P_{\omega_{\rm d}}^{(j)}$
for the TLS frequencies $\omega_{\rm d}^{(j)}$. The prefactors $P_{1,{\rm p}}$ and $P_{2,{\rm p}}$ denote the prior probabilities for the presence of at least one and two defects, respectively,
and the denominator ensures the conservation of probability. Note that if $P_{1,{\rm p}}+P_{2,{\rm p}}=0$, the value of $\omega_{\rm q}$ is irrelevant since by Eqs.~(\ref{eq:groundstate_oc}) and (\ref{eq:incohsum}), the qubit excited state probability does not depend on frequency. Moreover, $r$ is a uniform random variable on the interval $(0,1]$ and $\hat{T}_1$ denotes the mean over the posterior $\hat{T}_1=\int T_1 P({\boldsymbol x}|D)d{\boldsymbol x}$, with $\boldsymbol{x}=(g_1,g_2,\omega_{\rm d}^{(1)},\omega_{\rm d}^{(2)},T_{2,{\rm d}}^{(1)},T_{2,{\rm d}}^{(2)},T_{1,{\rm q}})$
encoding all the unknown model parameters. The relaxation time $T_1$ at different parameter values is obtained through Eq.~(\ref{eq:incohsum}).
In Fig.~(\ref{fig:swap}), the filled circles illustrate a simulated sequence of adaptive measurement settings, generated by the policy (\ref{eq:P_omega_q}), in the $\omega_{\rm q}$-$t$ plane. 
While the settings are initially uniformly distributed on the frequency interval, they later concentrate on the most relevant frequency ranges close to $\omega_{\rm d}^{(1)}$ and $\omega_{\rm d}^{(2)}$.
This would not be possible with measurement settings chosen prior to data collection which tends to make such measurements less efficient.
As directly implied by Eq.~(\ref{eq:P_omega_q}), the width of the range of waiting times $t$ is proportional to the expected $T_1$ at a given frequency. Without incorporating the current knowledge about 
$T_1$, one would tend to choose $t$ either too short or too long, rendering the measurement outcomes less sensitive to model parameters and leading to less accurate estimates.
\section{Initial prior}
\label{sec:iprior}
To evaluate the performance of our policy, we have applied it to large numbers of simulated samples with the true number of defects $n_{\rm d}$ equal to 0, 1, and 2. The parameters characterizing a sample are chosen uniformly in random such that $g_1,g_2\in [0.34,0.46]$ MHz, $\omega_{\rm d}^{(1)},\omega_{\rm d}^{(2)}\in [\omega_0-60,\omega_0+60]$ MHz, $T_{2,{\rm d}}^{(1)},T_{2,{\rm d}}^{(2)}\in [50,100]$ ns, and $T_{1,{\rm q}}\in [30,44]\ \mu s$. Note that the frequency $\omega_0$ denoting the mean of the interval does not play a role in the following discussion. The time interval for $T_{2,{\rm d}}^{(1)},T_{2,{\rm d}}^{(2)}$ is chosen to be consistent with the experimental observations in \cite{martinis05,shalibo10,barends13}, and the interval for $T_{1,\rm q}$ is chosen to correspond typical values for Xmon qubits \cite{barends13}. The parameter region for the coupling strengths and the defect frequencies is such that it can contain 2 or a smaller number of defects in experiments with transmons \cite{barends13}.

In the initial prior, the maximum number of defects is assumed to be 2 ($n_{\rm d}\le 2$), so that their number can be either 0, 1, or 2, each with the probability $\frac{1}{3}$. Given that the probabilistic information on the different quantities presented in the previous paragraph is everything that is known about a particular sample before data collection, the initial prior is chosen to be consistent with the probability distributions described above which uniquely determines the initial prior. We emphasize that we use the same initial prior for all the figures below, i.e., neither the true number of defects nor the values of any model parameters (except for $\hbar$) are assumed to be {\it a priori} known precisely. We label the defects such that the one with the higher (lower) $g$ is referred with the subscript 1 (2). In Figs.~\ref{fig:defects2} and \ref{fig:defects01}, all the curves are obtained by considering \mbox{10 000} simulated samples and in Fig.~\ref{fig:abspres} through 100 000 samples. 

\section{Sequential Monte Carlo method for simultaneous adaptive model selection and parameter estimation}
\label{sec:extended}
We implement the adaptive Bayesian inference scheme numerically through the sequential Monte Carlo (SMC) method \cite{west93,gordon93,liu01} that has, due to its computational efficiency, recently gained popularity in the context of quantum measurements \cite{huszar12,granade12,ferrie13,kravtsov13,wiebe14a,wiebe14b,stenberg14,granade15,stenberg15,stenberg16,struchalin16}. It approximates a continuous probability distribution $P({\boldsymbol x}|D)$ through $N_{\rm p}$ moving grid points or ``particles'' that are characterized
by their locations $\mathcal{S}=\{{\boldsymbol x_i}\}_{i=1}^{N_{\rm p}}$ in the parameter space and their relative probabilities, called weights, $\{w_i\}_{i=1}^{N_{\rm p}}$, that satisfy the normalization condition 
\begin{equation}
\sum_{i=1}^{N_{\rm p}}w_i = 1
\label{eq:normalization}
\end{equation}
due to conservation of probability. We perform the computations with $N_{\rm p}=40\ 000$. Once new data $d_{n+1}$ has been obtained, Bayes' theorem (\ref{eq:iter_bayes}) implies that the likelihood function may be used to update 
a weight $w_i$ to its new value $w_i'$
\begin{equation}
w_i'\sim P(d_{n+1}|{\boldsymbol x}_n)w_i.
\end{equation}
To cast this proportionality as an equality, an overall prefactor that is determined by the normalization condition (\ref{eq:normalization}) has to be assigned on the right-hand side.

We note that the SMC scheme described, e.g., in \cite{west93,gordon93,liu01,huszar12,granade12,ferrie13,kravtsov13,wiebe14a,wiebe14b,stenberg14,granade15,stenberg15,stenberg16,struchalin16} assumes a fixed, known, number of parameters to be estimated, and does not therefore adequately apply to our problem where we intend to characterize an {\it a priori} unknown number of defects in a sample. To point where the problems would arise, let us briefly review the role of the so-called resampling of the particles in the algorithm. Resampling is an element of the SMC scheme but does not constitute the whole algorithm; for the complete presentation of the algorithm, we refer to \cite{liu00,granade12,stenberg16}. Resampling imposes on the particles artificial dynamics whose purpose is to ``smoothen'' the discrete representation of the probability distribution $P({\boldsymbol x}|d_{n})$ and to collect the particles in the regions where the probability density is the highest, thus mitigating a limitation in accuracy that a fixed grid would cause. By construction, the artificial dynamics is constrained so that the expected mean and the covariances of the probability distibution are conserved.

Resampling assigns a new particle position in three steps denoted by (i)-(iii) below. These steps are implemented for every particle, i.e., $N_{\rm p}$ times for each update of the posterior. First, in step (i), among the current particle positions $\mathcal{S}$, one is chosen randomly following the discrete probability distribution $\{w_i\}_{i=1}^{N_{\rm p}}$. Let us assume this is the $n$th choice we make in the current update of the posterior and denote the location by ${\boldsymbol x}_n$. In step (ii), the position of this particle is shifted slightly to define the mean ${\boldsymbol \mu}_{n}$ for the $n$th resampling distribution
\begin{equation}
{\boldsymbol \mu}_{n}=a{\boldsymbol x}_n+(1-a)\hat{\boldsymbol x}.
\label{eq:shift}
\end{equation}
Here, the approximate mean $\hat{\boldsymbol x}$ of the particle locations, i.e., the current estimate of ${\boldsymbol x}$, is obtained through $\hat{\boldsymbol x}=\sum_{i=1}^{N_{\rm p}}w_i{\boldsymbol x_i}$, and $0\le a\le 1$ is a parameter that determines the location of ${\boldsymbol \mu}_{n}$ on a line connecting ${\boldsymbol x}_n$ and $\hat{\boldsymbol x}$. For the problem at hand, we obtain the best estimates with approximately $a=0.995$. The purpose of the second step is to ``compress'' the probability distribution to counteract the third step that increases the covariances. Finally, in step (iii), the new particle position is assigned by sampling from a (generally multidimensional) normal distribution 
\begin{equation}
{\boldsymbol x}_n'\sim \mathcal{N}[{\boldsymbol \mu}_{n},{\boldsymbol \Sigma}]
\label{eq:resampling}
\end{equation}
with the mean ${\boldsymbol \mu}_{n}$ and the covariance matrix ${\boldsymbol \Sigma}$. The matrix ${\boldsymbol \Sigma}$ is defined by
\begin{equation}
{\boldsymbol \Sigma}=(1-a^2){\rm Cov}[{\boldsymbol x}],
\label{eq:sigmadef}
\end{equation}
where ${\rm Cov}[{\boldsymbol x}]$ is the covariance matrix calculated over the particle positions $\mathcal{S}$. In Eq.~(\ref{eq:shift}), also ${\boldsymbol \mu}_{n}$ is a random variable since its value depends on the random variable ${\boldsymbol x}_n$, and the role of the prefactor $(1-a^2)\le1$ in Eq.~(\ref{eq:sigmadef}) is to ensure that the covariances are conserved in resampling.

In our example, we model a sample with two defects ($n_{\rm d}=2$) by a vector of the form $\boldsymbol{x}^{(2)}=(g_1,g_2,\omega_{\rm d}^{(1)},\omega_{\rm d}^{(2)},T_{2,{\rm d}}^{(1)},T_{2,{\rm d}}^{(2)},T_{1,{\rm q}})\in \mathbb{R}^7$. Since it is computationally convenient that all the particles are represented by vectors of the same length, we represent particles with $n_{\rm d}=1$ ($n_{\rm d}=0$) by vectors of the form $\boldsymbol{x}^{(1)}=(g_1,0,\omega_{\rm d}^{(1)},0,T_{2,{\rm d}}^{(1)},0,T_{1,{\rm q}})$ [$\boldsymbol{x}^{(0)}=(0,0,0,0,0,0,T_{1,{\rm q}})$]. This yields the correct observables. We also introduce the following shorthand notation for the sets of vectors describing samples with 0,1, and 2 defects
\begin{align}
&\mathcal{S}^{(0)}=\{{\boldsymbol x}_n^{(0)}\in \mathcal{S}|g_{1,n}=0,g_{2,n}=0\},\nonumber\\
&\mathcal{S}^{(1)}=\{{\boldsymbol x}_n^{(1)}\in \mathcal{S}|g_{1,n}\neq 0,g_{2,n}=0\},\nonumber\\
&\mathcal{S}^{(2)}=\{{\boldsymbol x}_n^{(2)}\in \mathcal{S}|g_{1,n}\neq 0,g_{2,n}\neq 0\}.
\label{eq:Ssets}
\end{align}
Since we assume that the number of defects can only have the values $n_{\rm d}=0,1,2$, we trivially have $\mathcal{S}=\bigcup_{k=0}^{2}\mathcal{S}^{(k)}$ and the intersection of the different sets is empty $\mathcal{S}^{(k)}\bigcap \mathcal{S}^{(j)}=\emptyset$, $k\neq j$.

Let us now consider ${\boldsymbol x}^{(k)}_n$ ($k=0,1$) that initially belongs to $\mathcal{S}^{(0)}$ or $\mathcal{S}^{(1)}$. We note that in resampling, shifting of the vector components in step (ii) through Eq.~(\ref{eq:shift}) would generally move ${\boldsymbol x}^{(k)}_n$ to $\mathcal{S}^{(2)}$. This is clearly an incorrect outcome since steps (ii) and (iii) are merely an unphysical computational procedure. They should not generally immediately change $n_{\rm d}$ to its maximum value for all the particles, but this is what would happen 
if the standard formulation of SMC would be applied without any modifications. A logical conclusion is that in step (ii), the components $g_2,\omega_{\rm d}^{(2)},T_{2,{\rm r}}^{(2)}$ (components $g_1,g_2,\omega_{\rm d}^{(1)},\omega_{\rm d}^{(2)},T_{2,{\rm d}}^{(1)},T_{2,{\rm d}}^{(2)}$) of ${\boldsymbol x}^{(1)}_n\in\mathcal{S}^{(1)}$ (${\boldsymbol x}^{(0)}_n\in\mathcal{S}^{(0)}$) should not be shifted at all.

But if the SMC algorithm is modified this way, the rest of the scheme has to be modified too, in order to conserve the covariances of the probability distribution. 
We find that covariances may be conserved by modifying resampling as follows, treating probability distributions for different numbers of particles separately. 
In step (i) of the resampling, we pick randomly ${\boldsymbol x}_n\in \mathcal{S}$ as in the standard scheme. As explained above, ${\boldsymbol x}_n\in \mathcal{S}^{(k)}$
with exactly one $k$ such that we may denote ${\boldsymbol x}_n={\boldsymbol x}_n^{(k)}$. In step (ii), we replace Eq.~(\ref{eq:shift}) by the equation
 ${\boldsymbol \mu}_{n}^{(k)}=a{\boldsymbol x}^{(k)}_n+(1-a)\hat{\boldsymbol x}^{(k)}$, where the superscript in ${\boldsymbol \mu}_{n}^{(k)}$ and $\hat{\boldsymbol x}^{(k)}$
 signifies that they are evaluated over $\mathcal{S}^{(k)}$ rather than $\mathcal{S}$. In (iii), we replace Eq.~(\ref{eq:resampling}) by ${{{\boldsymbol x}'}_n^{(k)}}\sim \mathcal{N}[{\boldsymbol \mu}_{n}^{(k)},{\boldsymbol \Sigma}^{(k)}]$, with ${\boldsymbol \Sigma}^{(k)}=(1-a^2){\rm Cov}[{\boldsymbol x^{(k)}}]$ where ${\rm Cov}[{\boldsymbol x^{(k)}}]$ is the covariance matrix evaluated over $\mathcal{S}^{(k)}$ rather than $\mathcal{S}$. When the SMC algorithm is modified this way, steps (ii) and (iii) in resampling conserve the number of defects $n_{\rm d}$ as well as the expected means and covariances for each $\mathcal{S}^{(k)}$ with $k=0,1,2$ separately. Since the expected means and the covariances are conserved for each $\mathcal{S}^{(k)}$ separately, they are also conserved for $\mathcal{S}$ which satisfies our goal. For the reasons presented above, the computations with an unmodified resampling algorithm would be marred by severe numerical errors, but the modified resampling scheme mitigates these issues. To summarize, one can say that the idea of our generalized SMC scheme is to implement the standard scheme for different sets $\mathcal{S}^{(k)}$ separately. 

Within our generalized SMC scheme, the probabilities $P_{k,{\rm p}}$ ($P_{k,{\rm a}}$) for the presence of at least $k$ defects (less than $k$ defects) in the sample may be approximated after counting the particles in $\mathcal{S}^{(k)}$, $k=0,1,2$. For instance, let us denote the set of indices of the particles in $\mathcal{S}^{(1)}$ by $\mathcal{I}^{(1)}$ such that 
$\mathcal{S}^{(1)}=\{x_i|i\in \mathcal{I}^{(1)}\}$. We can then approximate $P_{1,{\rm p}}=\sum_{i\in \mathcal{I}^{(1)}\bigcup\mathcal{I}^{(2)}}w_i$. The probabilities $P_{k,{\rm a}}$ and $P_{k,{\rm p}}$ are discussed in the context of Fig.~\ref{fig:abspres}.

The computations presented within this article apply the generalized SMC method described above and we carry them out with a computer program using Python and Fortan programming languages and NumPy \cite{ascher01} and SciPy \cite{jones01} software packages. 

\section{Results}
\label{sec:results}
\begin{figure}[h!]
\includegraphics[width=0.4\textwidth]{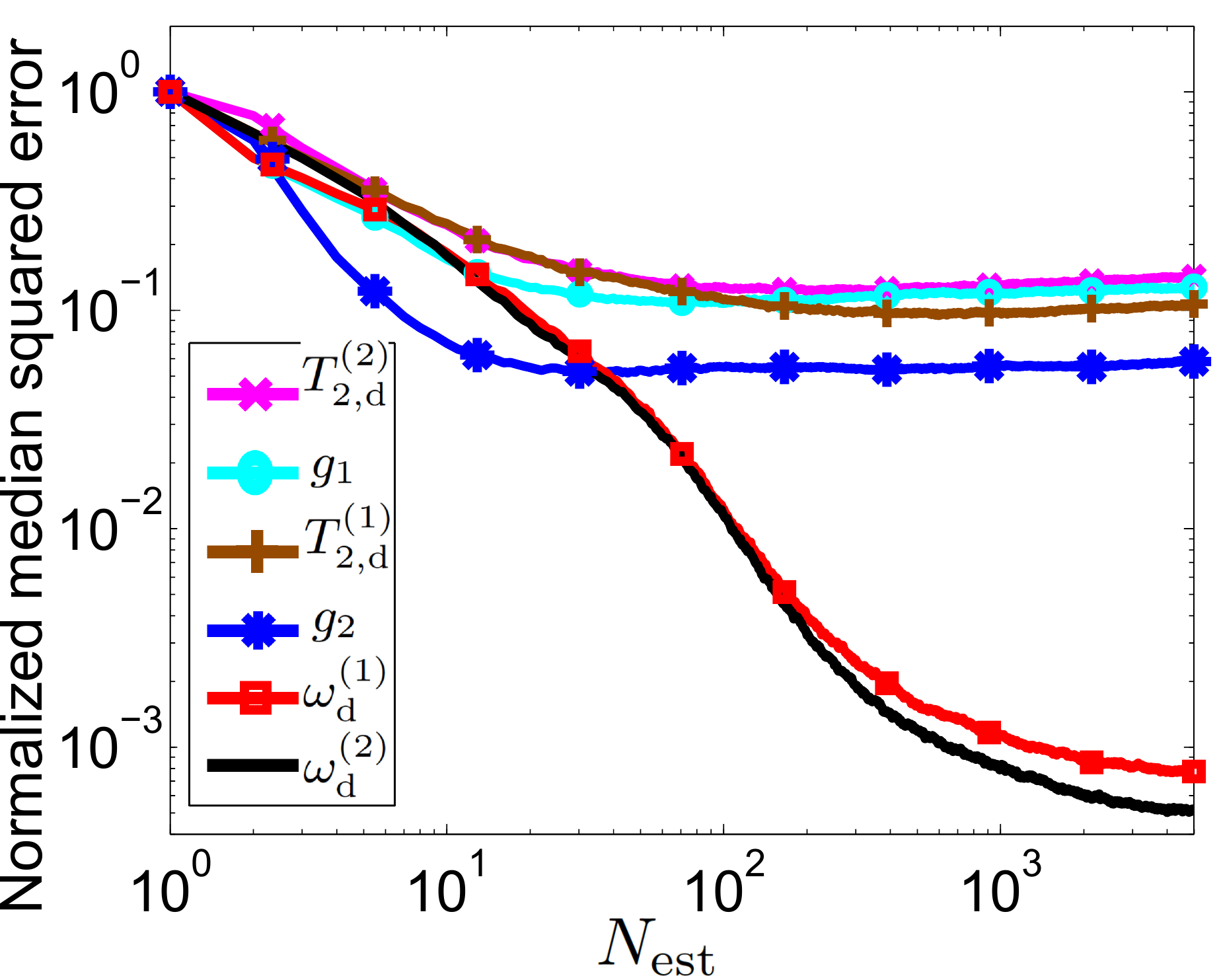}
\caption{Accuracy of parameter estimation in the presence of two defects in a sample ($n_{\rm d}=2$). Normalized median squared error as a function of the number of estimates $N_{\rm est}$ for different parameters of the model described in Sec.~\ref{sec:model}: $g_1$ (cyan with circles), $g_2$ (blue with asterisks), 
$\omega_{\rm d}^{(1)}$ (red with squares), $\omega_{\rm d}^{(2)}$ (black solid), $T_{2,{\rm d}}^{(1)}$ (brown with `+' symbols), and $T_{2,{\rm d}}^{(2)}$ (magenta with crosses).}
\label{fig:defects2} \end{figure}

Figure \ref{fig:defects2} presents the results for samples with exactly two defects ($n_{\rm d}=2$). Here, normalized median squared errors are plotted as a function of the number of 
estimates $N_{\rm est}$ obtained (one for the initial prior and one for each measurement setting). For each measurement setting, the measurement is repeated $M_{\rm r}=200$ times, such that the total number of measurement shots is $M=M_{\rm r} (N_{\rm est}-1)$. We find that the errors for $\omega_{\rm d}^{(1)}$ and $\omega_{\rm d}^{(2)}$ decrease monotonically over the range of $N_{\rm est}$ studied, but the errors for $g_1$, $g_2$, $T_{2,{\rm d}}^{(1)}$,$T_{2,{\rm d}}^{(2)}$ saturate at $N_{\rm est}$ of the order of $\sim 10^2$.  
Typical observable relaxation times $T_1$ of a qubit (an example may be inferred from Fig.~\ref{fig:swap}) are at least an order of magnitude shorter than typical values of $T_{1,{\rm q}}$ (cf. Sec.~\ref{sec:iprior}). Due to relative weakness of the corresponding signal, we are therefore unable to considerably improve the estimates of $T_{1,{\rm q}}$ from their initial values (not shown in the figure), but the uncertainty in $T_{1,{\rm q}}$ still allows us to improve the estimates of $\omega_{\rm d}^{(1)}$ and $\omega_{\rm d}^{(2)}$ which are  experimentally the most relevant quantities.

\begin{figure}[h!]
\includegraphics[width=0.4\textwidth]{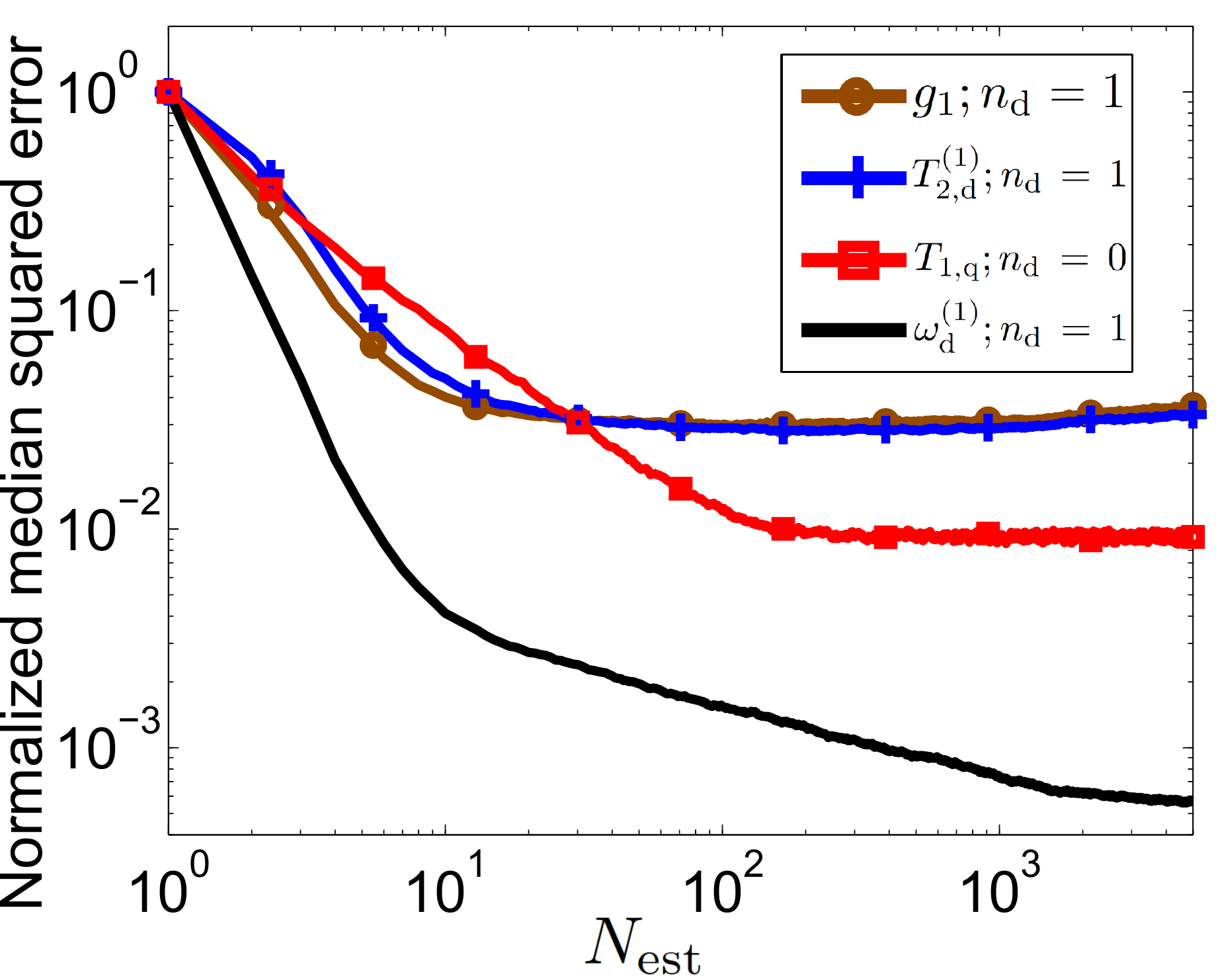}
\caption{Accuracy of parameter estimation in the absence of any defects in a sample ($n_{\rm d}=0$) and in the presence of a single defect ($n_{\rm d}=1$). Normalized median squared error as a function of the number of estimates $N_{\rm est}$ for different parameters of the model described in Sec.~\ref{sec:model}: $g_1$ (brown with circles), $\omega_{\rm d}^{(1)}$ (black solid), $T_{2,{\rm d}}^{(1)}$ (blue with `+' symbols), and $T_{1,{\rm q}}$ (red with squares).}
\label{fig:defects01}
\end{figure}

The samples with a single defect ($n_{\rm d}=1$) and with no defects ($n_{\rm d}=0$) are described in Fig.~\ref{fig:defects01}. A single defect can be characterized by the quantities $g_1$, $\omega_{\rm d}^{(1)}$, and $T_{2,{\rm d}}^{(1)}$ and the normalized median squared errors of the estimates of these quantities are here plotted as a function of $N_{\rm est}$. In the absence of any defects, the qubit relaxation is determined by the frequency-independent $T_{1,{\rm q}}$. At $N_{\rm est}\sim 10^2$, its median squared error is decreased to the level that with the given initial uncertainty corresponds to the uncertainty of approximately $\sim 1\mu s$, but with larger $N_{\rm est}$ the error saturates due to issues that we attribute to numerical accuracy. In the presence of a defect, similarly to Fig.~\ref{fig:defects2}, due to weakness of the signal, we are not able to considerably improve the estimates of $T_{1,{\rm q}}$ (not shown in the figure), but the remaining uncertainty in $T_{1,{\rm q}}$ nevertheless allows to decrease the error of $\omega_{\rm d}^{(1)}$.

\begin{figure}[h!]
\includegraphics[width=0.4\textwidth]{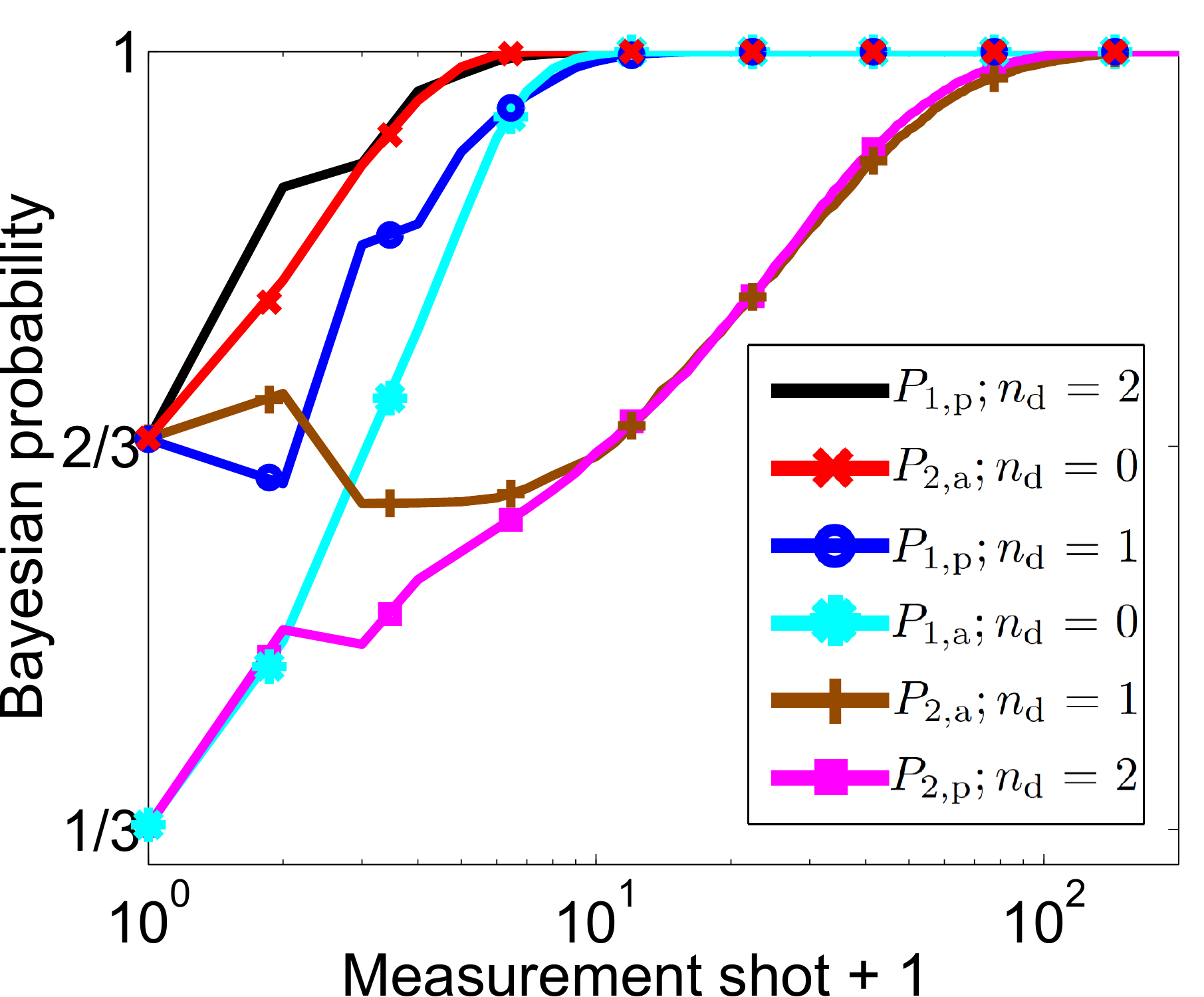}
\caption{Model selection in the presence of an unknown number of defects in a sample, cf. Eq.~(\ref{eq:lindblad}). Different curves exhibit the medians of the Bayesian probabilities as a function of the number of measurement shots. The Bayesian probabilities are for the presence of at least a single defect ($P_{1,{\rm p}}$), for the presence of a second defect ($P_{2,{\rm p}}$), for the absence of any defects ($P_{1,{\rm a}}$), and for the absence of a second defect ($P_{2,{\rm a}}$). The true number of defects is denoted by $n_{\rm d}=0,1,2$.}
\label{fig:abspres}
\end{figure}

Above, we have not yet shown how quickly the policy learns about the possible absence of defects in the samples. We find that when $M_{\rm r}=200$ as above, a single measurement settings is sufficient to find the absence or presence of the defects (not shown in the figures). To more precisely quantify this efficiency, we now reduce the number of repetitions of the measurements, and thereby the information gained, at a single setting and set $M_{\rm r}=1$ such that the number of measurement shots is simply $M=M_{\rm r} (N_{\rm est}-1)=N_{\rm est}-1$. In Fig.~\ref{fig:abspres}, we plot the median Bayesian probabilities for the three diffent pieces of knowledge describing the absence of defects. These are the probability for the absence of any defects $P_{1,{\rm a}}$, when the true number of defects is $n_{\rm d}=0$ (cyan curve with asterisks), as well as the probabilities for the absence of a second defect $P_{2,{\rm a}}$, with $n_{\rm d}=1$ (brown with `+' symbols) and with $n_{\rm d}=0$ (red with crosses). The figure also exhibits median Bayesian probabilities quantifying three diffent (correct) beliefs in the presence of defects. These are the probability for the existence of at least a single defect $P_{1,{\rm p}}$, with $n_{\rm d}=1$ (blue with circles) and with $n_{\rm d}=2$ (black solid), and the probability for the existence of a second defect $P_{2,{\rm p}}$, when $n_{\rm d}=2$ (magenta). The initial values of the probabilities are determined by the initial prior described in Sec.~\ref{sec:iprior}. All the probabilities except $P_{2,{\rm a}}$ with $n_{\rm d}=1$ and $P_{2,{\rm p}}$ with $n_{\rm d}=2$ reach a level virtually indistinguishable from unity, i.e., certainty about a belief that is correct, after $M\sim$ 10 measurement shots whereas the latter reach such a value after $M\sim 10^2$ shots. Note that here, the distance to unity can be interpreted as error in model selection since it represents the Bayesian probability of the all the other models except the correct one. The initial prior and Bayes' theorem (\ref{eq:iter_bayes}) imply that the possible number of defects allowed in any posterior can not exceed 2. Comparing the results of Fig.~\ref{fig:abspres} with those in Figs.~\ref{fig:defects2} and \ref{fig:defects01} shows that here, model selection requires considerably less measurements than parameter estimation.

\section{Conclusion}
We have generalized the concept of adaptive measurements to account for simultaneous model selection and parameter estimation. We implemented a policy built on an adaptive Bayesian scheme using a generalization of sequential Monte Carlo method. Since global optimization through utility \cite{sivia06} maximization would generally be numerically challenging, many recent works making use of Bayesian inference have adopted either greedy algorithms \cite{berry00,fischer00,berry01,sergeevich11,huszar12,ferrie13,kravtsov13,struchalin16} that optimize a certain utility function assuming the next measurement is the last one, or different heuristics that choose measurement settings adaptively through a probability distribution \cite{wiebe14a,wiebe14b,stenberg14,stenberg15,stenberg16}. Our work belongs to the latter category. 

Assuming the model is known and that only its parameter values are uncertain, it should be noted that the difficulty of parameter estimation depends on the model. In certain cases it is possible to drastically outperform nonadaptive measurements and, e.g., to improve the accuracy of the estimate exponentially as a function of the number of measurement shots
 \cite{sergeevich11,ferrie13,wiebe14a,wiebe14b,stenberg14}. Basically, this is because in the systems considered in \cite{sergeevich11,ferrie13,wiebe14a,wiebe14b,stenberg14}, small changes in the coupling strength may correspond to a large change in the observable $P_{\rm e}$ when $t$ is long. However, in the system studied within this paper (with the parameters in the region discussed in Sec.~\ref{sec:iprior}), long $t$ leads to loss of the signal due to rapid relaxation of the qubit which is why small changes in $g_j$ and $T_{2,{\rm d}}^{(j)}$ always correspond to small differences in $P_{\rm e}$. Hence inferring $g_j$ and $T_{2,{\rm d}}^{(j)}$ is difficult. Estimation of $T_{1,{\rm q}}$ is challenging because the competing contribution of $\Gamma_{1,{\rm d}}^{(j)}$ (that depends on $g_j$ and $T_{2,{\rm d}}^{(j)}$) on $P_{\rm e}$ is much stronger, cf. Eqs.~(\ref{eq:groundstate_oc}) and (\ref{eq:incohsum}) and Sec.~\ref{sec:iprior}. Furthermore, since the volume of the parameter space increases exponentially with its dimension, this makes the set of particles sparser. This ``curse of dimensionality'' decreases the effective sample size $N_{\rm ess}=\frac{1}{\sum_{i}w_i^2}$, making it smaller than, e.g., in the considerations of \cite{stenberg14,stenberg15,stenberg16}, which may increase the numerical error. We attribute the observed floors in accuracy in Figs.~\ref{fig:defects2} and \ref{fig:defects01} to competition between information obtained through measurements and accumulated numerical error. Despite these challenges, we are able to decrease the squared error of $\omega_{\rm d}^{(j)}$ by a factor of $\sim 10^3$ by making only $\sim 10^3$ updates in the measurement setting. We emphasize that $\omega_{\rm d}^{(j)}$ are the quantities that are the most relevant, e.g., to protect a qubit from unwanted decoherence since avoiding parking the qubit at these frequencies makes its coherence time longer. 
 
Even though in this paper we have considered characterization of incoherent TLSs as an example, the idea of simultaneous adaptive model selection and parameter estimation is applicable to a larger class of problems. For instance, once the required likelihood functions have been derived, it is straightforward to generalize the work in \cite{stenberg14,stenberg16} for an unknown number of coherent TLSs. 

For a larger number of TLSs in the frequency interval, we note that when their frequencies $\omega_i$ are sufficiently far from each others to neglect the overlaps in their swap spectra, a simple approximation holds. Denoting the mean squared error of a TLS frequency $\omega_i$ by $\mathcal{E}^2_{\omega_i}$, we expect that $\frac{1}{n_{\rm d}}\sum_{i=1}^{n_{\rm d}}\mathcal{E}^2_{\omega_i}(Mn_{\rm d})|_{n_{\rm d}}\approx\mathcal{E}^2_{\omega_1}(M)|_{n_{\rm d=1}}$, i.e., the mean squared error averaged over $n_{\rm d}$ defects after $Mn_{\rm d}$ measurement shots is approximately equal to $\mathcal{E}^2_{\omega_1}$ after $M$ shots in the presence of a single defect because in the former case the measurements have to be allocated among a larger number ($n_{\rm d}$) of different TLSs. 
In a more realistic scenario, one has to take into account the overlaps in the swap spectra which is difficult without doing the full numerical calculation.

Another topic for further research is the influence of imperfections in the readout. Such errors are best tolerated when first quantified and then incorporated in the likelihood function. For instance, when the probability of misidentifying a qubit ground state as an excited state or vice versa equals $\gamma$ one has to replace Eq.~(\ref{eq:groundstate_oc}) by $P_{\rm e}=(1-2\gamma)\exp[-t/T_1]+\gamma$. As discussed in Sec.~\ref{sec:results}, selecting the correct model usually requires much less data than achieving a reasonable degree of accuracy in parameter estimation, i.e., the correct model is usually found in a very early stage of the parameter estimation. On the other hand, adaptive Bayesian inference scheme combined with SMC approximation has been shown to be robust against substantial amount of readout errors for parameter estimation of different fixed models \cite{wiebe14a,wiebe14b,stenberg14,stenberg16} as well for estimation of different quantum states \cite{kravtsov13,stenberg15,struchalin16}. We thus expect that robustness against readout errors can also quite generally be achieved for simultaneous adaptive model selection and parameter estimation.

In conclusion, assuming both the model and all its parameters are initially uncertain, we formulated system characterizion in a unified Bayesian framework and delivered numerical tools required for solving a concrete problem. We applied our method for efficient detection and characterization of a sparse bath of incoherent spurious two-level systems, a leading decoherence mechanism in the state-of-the-art superconducting qubits.
\section*{ACKNOWLEDGEMENTS}
We acknowledge O. K\"ohn for discussions. This work was supported by the European Union through ScaleQIT.
\appendix
\section{Lindblad operator}
\label{app:lindblad}
When the TLSs interact incoherently with the qubit, Markovian decoherence can be described by the Lindblad operator \cite{nielsen00}
\begin{equation}
\mathcal{D}[\hat{C}_{k}^{(j)}]\hat{\rho}=\hat{C}_{k}^{(j)}\hat{\rho} \hat{C}_{k}^{(j)\dag}-\frac{1}{2}
\{\hat{C}_{k}^{(j)\dag} \hat{C}_{k}^{(j)},\hat{\rho}\},
\end{equation}
where the terms 
\begin{align}
&\hat{C}_{1}^{(j)}=\hat{a}\sqrt{\Gamma_{1,\rm{q}}^{(j)}},\quad \hat{C}_{2}^{(j)}=\hat{a}^{\dag}\hat{a}\sqrt{2\Gamma_{\phi,\rm{q}}^{(j)}},\nonumber\\
&\hat{C}_{3}^{(j)}=\hat{\sigma}_{-}\sqrt{\Gamma_{1,\rm{d}}^{(j)}},\quad \hat{C}_{4}^{(j)}=\hat{\sigma}_{+}\hat{\sigma}_{-}\sqrt{2\Gamma_{\phi,\rm{d}}^{(j)}},
\end{align}
denote the energy and phase relaxation rates of the qubit and the defect.


\begin{thebibliography}{58}
\expandafter\ifx\csname natexlab\endcsname\relax\def\natexlab#1{#1}\fi
\expandafter\ifx\csname bibnamefont\endcsname\relax
  \def\bibnamefont#1{#1}\fi
\expandafter\ifx\csname bibfnamefont\endcsname\relax
  \def\bibfnamefont#1{#1}\fi
\expandafter\ifx\csname citenamefont\endcsname\relax
  \def\citenamefont#1{#1}\fi
\expandafter\ifx\csname url\endcsname\relax
  \def\url#1{\texttt{#1}}\fi
\expandafter\ifx\csname urlprefix\endcsname\relax\def\urlprefix{URL }\fi
\providecommand{\bibinfo}[2]{#2}
\providecommand{\eprint}[2][]{\url{#2}}

\bibitem[{K. P. Burnham and D. R. Anderson, {\it Model Selection and Multimodel
  Inference: A Practical Information-Theoretic Approach} (Springer, New York,
  2002)()}]{burnhambook}
K. P. Burnham and D. R. Anderson, {\it Model Selection and Multimodel
  Inference: A Practical Information-Theoretic Approach} (Springer, New York,
  2002).

\bibitem[{\citenamefont{Akaike}(1974)}]{akaike74}
\bibinfo{author}{\bibfnamefont{H.}~\bibnamefont{Akaike}},
  \bibinfo{journal}{IEEE Trans. Autom. Control} \textbf{\bibinfo{volume}{19}},
  \bibinfo{pages}{716} (\bibinfo{year}{1974}).

\bibitem[{\citenamefont{Schwarz}(1978)}]{schwarz78}
\bibinfo{author}{\bibfnamefont{G.}~\bibnamefont{Schwarz}},
  \bibinfo{journal}{Ann. Statist.} \textbf{\bibinfo{volume}{6}},
  \bibinfo{pages}{461} (\bibinfo{year}{1978}).

\bibitem[{D. Sivia and J. Skilling, {\it Data Analysis: A Bayesian Tutorial}
  (Oxford University Press, New York, 2006).()}]{sivia06}
D. Sivia and J. Skilling, {\it Data Analysis: A Bayesian Tutorial} (Oxford
  University Press, New York, 2006).

\bibitem[{\citenamefont{{Y. Nakamura} et~al.}(1999)\citenamefont{{Y. Nakamura},
  {Yu. A. Pashkin}, and {J. S. Tsai}}}]{nakamura99}
\bibinfo{author}{\bibnamefont{{Y. Nakamura}}},
  \bibinfo{author}{\bibnamefont{{Yu. A. Pashkin}}}, \bibnamefont{and}
  \bibinfo{author}{\bibnamefont{{J. S. Tsai}}}, \bibinfo{journal}{Nature
  (London)} \textbf{\bibinfo{volume}{398}}, \bibinfo{pages}{786}
  (\bibinfo{year}{1999}).

\bibitem[{\citenamefont{{J. Koch} et~al.}(2007)\citenamefont{{J. Koch}, {T. M.
  Yu}, {J. Gambetta}, {A. A. Houck}, {D. I. Schuster}, {J. Majer}, {A. Blais},
  {M. H. Devoret}, {S. M. Girvin}, and {R. J. Schoelkopf}}}]{koch07}
\bibinfo{author}{\bibnamefont{{J. Koch}}}, \bibinfo{author}{\bibnamefont{{T. M.
  Yu}}}, \bibinfo{author}{\bibnamefont{{J. Gambetta}}},
  \bibinfo{author}{\bibnamefont{{A. A. Houck}}},
  \bibinfo{author}{\bibnamefont{{D. I. Schuster}}},
  \bibinfo{author}{\bibnamefont{{J. Majer}}}, \bibinfo{author}{\bibnamefont{{A.
  Blais}}}, \bibinfo{author}{\bibnamefont{{M. H. Devoret}}},
  \bibinfo{author}{\bibnamefont{{S. M. Girvin}}}, \bibnamefont{and}
  \bibinfo{author}{\bibnamefont{{R. J. Schoelkopf}}}, \bibinfo{journal}{Phys.
  Rev. A} \textbf{\bibinfo{volume}{76}}, \bibinfo{pages}{042319}
  (\bibinfo{year}{2007}).

\bibitem[{\citenamefont{{M. H. Devoret} and {R. J.
  Schoelkopf}}(2013)}]{devoret13}
\bibinfo{author}{\bibnamefont{{M. H. Devoret}}} \bibnamefont{and}
  \bibinfo{author}{\bibnamefont{{R. J. Schoelkopf}}},
  \bibinfo{journal}{Science} \textbf{\bibinfo{volume}{339}},
  \bibinfo{pages}{1169} (\bibinfo{year}{2013}).

\bibitem[{\citenamefont{{M. v. Schickfus} and {S.
  Hunklinger}}(1977)}]{schickfus77}
\bibinfo{author}{\bibnamefont{{M. v. Schickfus}}} \bibnamefont{and}
  \bibinfo{author}{\bibnamefont{{S. Hunklinger}}}, \bibinfo{journal}{Phys.
  Lett. A} \textbf{\bibinfo{volume}{64}}, \bibinfo{pages}{144}
  (\bibinfo{year}{1977}).

\bibitem[{\citenamefont{{J. M. Martinis} et~al.}(2005)\citenamefont{{J. M.
  Martinis}, {K. B. Cooper}, {R. McDermott}, {M. Steffen}, {M. Ansmann}, {K. D.
  Osborn}, {K. Cicak}, {S. Oh}, {D. P. Pappas}, {R. W. Simmonds}
  et~al.}}]{martinis05}
\bibinfo{author}{\bibnamefont{{J. M. Martinis}}},
  \bibinfo{author}{\bibnamefont{{K. B. Cooper}}},
  \bibinfo{author}{\bibnamefont{{R. McDermott}}},
  \bibinfo{author}{\bibnamefont{{M. Steffen}}},
  \bibinfo{author}{\bibnamefont{{M. Ansmann}}},
  \bibinfo{author}{\bibnamefont{{K. D. Osborn}}},
  \bibinfo{author}{\bibnamefont{{K. Cicak}}}, \bibinfo{author}{\bibnamefont{{S.
  Oh}}}, \bibinfo{author}{\bibnamefont{{D. P. Pappas}}},
  \bibinfo{author}{\bibnamefont{{R. W. Simmonds}}}, \bibnamefont{et~al.},
  \bibinfo{journal}{Phys. Rev. Lett.} \textbf{\bibinfo{volume}{95}},
  \bibinfo{pages}{210503} (\bibinfo{year}{2005}).

\bibitem[{\citenamefont{{R. Barends} et~al.}(2010)\citenamefont{{R. Barends},
  {N. Vercruyssen}, {A. Endo}, {P. J. Visser}, {T. Zijlstra}, {T. M. Klapwijk},
  {P. Diener}, {S. J. C. Yates}, and {J. J. A. Baselmans}}}]{barends10}
\bibinfo{author}{\bibnamefont{{R. Barends}}}, \bibinfo{author}{\bibnamefont{{N.
  Vercruyssen}}}, \bibinfo{author}{\bibnamefont{{A. Endo}}},
  \bibinfo{author}{\bibnamefont{{P. J. Visser}}},
  \bibinfo{author}{\bibnamefont{{T. Zijlstra}}},
  \bibinfo{author}{\bibnamefont{{T. M. Klapwijk}}},
  \bibinfo{author}{\bibnamefont{{P. Diener}}},
  \bibinfo{author}{\bibnamefont{{S. J. C. Yates}}}, \bibnamefont{and}
  \bibinfo{author}{\bibnamefont{{J. J. A. Baselmans}}}, \bibinfo{journal}{Appl.
  Phys. Lett.} \textbf{\bibinfo{volume}{97}}, \bibinfo{pages}{023508}
  (\bibinfo{year}{2010}).

\bibitem[{\citenamefont{{R. Barends {\it et al.}}}(2013)}]{barends13}
\bibinfo{author}{\bibnamefont{{R. Barends {\it et al.}}}},
  \bibinfo{journal}{Phys. Rev. Lett.} \textbf{\bibinfo{volume}{111}},
  \bibinfo{pages}{080502} (\bibinfo{year}{2013}).

\bibitem[{\citenamefont{{J. Gao} et~al.}(2008)\citenamefont{{J. Gao}, {M.
  Daal}, {A. Vayonakis}, {S. Kumar}, {J. Zmuidzinas}, {B. Sadoulet}, {B. A.
  Mazin}, {P. K. Day}, and {H. G. Leduc}}}]{gao08}
\bibinfo{author}{\bibnamefont{{J. Gao}}}, \bibinfo{author}{\bibnamefont{{M.
  Daal}}}, \bibinfo{author}{\bibnamefont{{A. Vayonakis}}},
  \bibinfo{author}{\bibnamefont{{S. Kumar}}}, \bibinfo{author}{\bibnamefont{{J.
  Zmuidzinas}}}, \bibinfo{author}{\bibnamefont{{B. Sadoulet}}},
  \bibinfo{author}{\bibnamefont{{B. A. Mazin}}},
  \bibinfo{author}{\bibnamefont{{P. K. Day}}}, \bibnamefont{and}
  \bibinfo{author}{\bibnamefont{{H. G. Leduc}}}, \bibinfo{journal}{Appl. Phys.
  Lett.} \textbf{\bibinfo{volume}{92}}, \bibinfo{pages}{152505}
  (\bibinfo{year}{2008}).

\bibitem[{\citenamefont{{H. Wang {\it et al.}}}(2009)}]{wang09}
\bibinfo{author}{\bibnamefont{{H. Wang {\it et al.}}}}, \bibinfo{journal}{Appl.
  Phys. Lett.} \textbf{\bibinfo{volume}{95}}, \bibinfo{pages}{233508}
  (\bibinfo{year}{2009}).

\bibitem[{\citenamefont{{J. Wenner {\it et al.}}}(2011)}]{wenner11}
\bibinfo{author}{\bibnamefont{{J. Wenner {\it et al.}}}},
  \bibinfo{journal}{Appl. Phys. Lett.} \textbf{\bibinfo{volume}{99}},
  \bibinfo{pages}{113513} (\bibinfo{year}{2011}).

\bibitem[{\citenamefont{{K. B. Cooper} et~al.}(2004)\citenamefont{{K. B.
  Cooper}, {M. Steffen}, {R. McDermott}, {R. W. Simmonds}, {S. Oh}, {D. A.
  Hite}, {D. P. Pappas}, and {J. M. Martinis}}}]{cooper04}
\bibinfo{author}{\bibnamefont{{K. B. Cooper}}},
  \bibinfo{author}{\bibnamefont{{M. Steffen}}},
  \bibinfo{author}{\bibnamefont{{R. McDermott}}},
  \bibinfo{author}{\bibnamefont{{R. W. Simmonds}}},
  \bibinfo{author}{\bibnamefont{{S. Oh}}}, \bibinfo{author}{\bibnamefont{{D. A.
  Hite}}}, \bibinfo{author}{\bibnamefont{{D. P. Pappas}}}, \bibnamefont{and}
  \bibinfo{author}{\bibnamefont{{J. M. Martinis}}}, \bibinfo{journal}{Phys.
  Rev. Lett.} \textbf{\bibinfo{volume}{93}}, \bibinfo{pages}{180401}
  (\bibinfo{year}{2004}).

\bibitem[{\citenamefont{{Y. Shalibo} et~al.}(2010)\citenamefont{{Y. Shalibo},
  {Y. Rofe}, {D. Shwa}, {F. Zeides}, {M. Neeley}, {J. M. Martinis}, and {N.
  Katz}}}]{shalibo10}
\bibinfo{author}{\bibnamefont{{Y. Shalibo}}}, \bibinfo{author}{\bibnamefont{{Y.
  Rofe}}}, \bibinfo{author}{\bibnamefont{{D. Shwa}}},
  \bibinfo{author}{\bibnamefont{{F. Zeides}}},
  \bibinfo{author}{\bibnamefont{{M. Neeley}}},
  \bibinfo{author}{\bibnamefont{{J. M. Martinis}}}, \bibnamefont{and}
  \bibinfo{author}{\bibnamefont{{N. Katz}}}, \bibinfo{journal}{Phys. Rev.
  Lett.} \textbf{\bibinfo{volume}{105}}, \bibinfo{pages}{177001}
  (\bibinfo{year}{2010}).

\bibitem[{\citenamefont{{T. Palom\"aki {\it et al.}}}(2010)}]{palomaki10}
\bibinfo{author}{\bibnamefont{{T. Palom\"aki {\it et al.}}}},
  \bibinfo{journal}{Phys. Rev. B} \textbf{\bibinfo{volume}{81}},
  \bibinfo{pages}{144503} (\bibinfo{year}{2010}).

\bibitem[{\citenamefont{{J. Lisenfeld} et~al.}(2010)\citenamefont{{J.
  Lisenfeld}, {C. M\"uller}, {J. H. Cole}, {P. Bushev}, {A. Lukashenko}, {A.
  Shnirman}, and {A. V. Ustinov}}}]{lisenfeld10}
\bibinfo{author}{\bibnamefont{{J. Lisenfeld}}},
  \bibinfo{author}{\bibnamefont{{C. M\"uller}}},
  \bibinfo{author}{\bibnamefont{{J. H. Cole}}},
  \bibinfo{author}{\bibnamefont{{P. Bushev}}},
  \bibinfo{author}{\bibnamefont{{A. Lukashenko}}},
  \bibinfo{author}{\bibnamefont{{A. Shnirman}}}, \bibnamefont{and}
  \bibinfo{author}{\bibnamefont{{A. V. Ustinov}}}, \bibinfo{journal}{Phys. Rev.
  Lett.} \textbf{\bibinfo{volume}{105}}, \bibinfo{pages}{230504}
  (\bibinfo{year}{2010}).

\bibitem[{\citenamefont{{G. J. Grabovskij} et~al.}(2012)\citenamefont{{G. J.
  Grabovskij}, {T. Peichl}, {J. Lisenfeld}, {G. Weiss}, and {A. V.
  Ustinov}}}]{grabovskij12}
\bibinfo{author}{\bibnamefont{{G. J. Grabovskij}}},
  \bibinfo{author}{\bibnamefont{{T. Peichl}}},
  \bibinfo{author}{\bibnamefont{{J. Lisenfeld}}},
  \bibinfo{author}{\bibnamefont{{G. Weiss}}}, \bibnamefont{and}
  \bibinfo{author}{\bibnamefont{{A. V. Ustinov}}}, \bibinfo{journal}{Science}
  \textbf{\bibinfo{volume}{338}}, \bibinfo{pages}{232} (\bibinfo{year}{2012}).

\bibitem[{\citenamefont{{B. Sarabi} et~al.}(2016)\citenamefont{{B. Sarabi}, {A.
  N. Ramanayaka}, {A. L. Burin}, {F. C. Wellstood}, and {K. D.
  Osborn}}}]{sarabi16}
\bibinfo{author}{\bibnamefont{{B. Sarabi}}}, \bibinfo{author}{\bibnamefont{{A.
  N. Ramanayaka}}}, \bibinfo{author}{\bibnamefont{{A. L. Burin}}},
  \bibinfo{author}{\bibnamefont{{F. C. Wellstood}}}, \bibnamefont{and}
  \bibinfo{author}{\bibnamefont{{K. D. Osborn}}}, \bibinfo{journal}{Phys. Rev.
  Lett.} \textbf{\bibinfo{volume}{116}}, \bibinfo{pages}{167002}
  (\bibinfo{year}{2016}).

\bibitem[{\citenamefont{Ferrie}(2014)}]{ferrie14b}
\bibinfo{author}{\bibfnamefont{C.}~\bibnamefont{Ferrie}}, \bibinfo{journal}{New
  J. Phys.} \textbf{\bibinfo{volume}{16}}, \bibinfo{pages}{093035}
  (\bibinfo{year}{2014}).

\bibitem[{\citenamefont{{J. Emerson} et~al.}(2005)\citenamefont{{J. Emerson},
  {R. Alicki}, and {K. Zyczkowski}}}]{emerson05}
\bibinfo{author}{\bibnamefont{{J. Emerson}}}, \bibinfo{author}{\bibnamefont{{R.
  Alicki}}}, \bibnamefont{and} \bibinfo{author}{\bibnamefont{{K. Zyczkowski}}},
  \bibinfo{journal}{J. Opt. B} \textbf{\bibinfo{volume}{7}},
  \bibinfo{pages}{S347} (\bibinfo{year}{2005}).

\bibitem[{\citenamefont{{E. Knill} et~al.}(2008)\citenamefont{{E. Knill}, {D.
  Leibfried}, {R. Reichle}, {J. Britton}, {R. B. Blakestad}, {J. D. Jost}, {C.
  Langer}, {R. Ozeri}, {S. Seidelin}, and {D. J. Wineland}}}]{knill08}
\bibinfo{author}{\bibnamefont{{E. Knill}}}, \bibinfo{author}{\bibnamefont{{D.
  Leibfried}}}, \bibinfo{author}{\bibnamefont{{R. Reichle}}},
  \bibinfo{author}{\bibnamefont{{J. Britton}}},
  \bibinfo{author}{\bibnamefont{{R. B. Blakestad}}},
  \bibinfo{author}{\bibnamefont{{J. D. Jost}}},
  \bibinfo{author}{\bibnamefont{{C. Langer}}},
  \bibinfo{author}{\bibnamefont{{R. Ozeri}}}, \bibinfo{author}{\bibnamefont{{S.
  Seidelin}}}, \bibnamefont{and} \bibinfo{author}{\bibnamefont{{D. J.
  Wineland}}}, \bibinfo{journal}{Phys. Rev. A} \textbf{\bibinfo{volume}{77}},
  \bibinfo{pages}{012307} (\bibinfo{year}{2008}).

\bibitem[{\citenamefont{{E. Magesan} et~al.}(2011)\citenamefont{{E. Magesan},
  {J. M. Gambetta}, and {J. Emerson}}}]{magesan11}
\bibinfo{author}{\bibnamefont{{E. Magesan}}}, \bibinfo{author}{\bibnamefont{{J.
  M. Gambetta}}}, \bibnamefont{and} \bibinfo{author}{\bibnamefont{{J.
  Emerson}}}, \bibinfo{journal}{Phys. Rev. Lett.}
  \textbf{\bibinfo{volume}{106}}, \bibinfo{pages}{180504}
  (\bibinfo{year}{2011}).

\bibitem[{\citenamefont{{E. Magesan {\it et al.}}}(2012)}]{magesan12}
\bibinfo{author}{\bibnamefont{{E. Magesan {\it et al.}}}},
  \bibinfo{journal}{Phys. Rev. Lett.} \textbf{\bibinfo{volume}{109}},
  \bibinfo{pages}{080505} (\bibinfo{year}{2012}).

\bibitem[{\citenamefont{{D. W. Berry} et~al.}(2009)\citenamefont{{D. W. Berry},
  {B. L. Higgins}, {S. D. Bartlett}, {M. W. Mitchell}, {G. J. Pryde}, and {H.
  M. Wiseman}}}]{berry09}
\bibinfo{author}{\bibnamefont{{D. W. Berry}}},
  \bibinfo{author}{\bibnamefont{{B. L. Higgins}}},
  \bibinfo{author}{\bibnamefont{{S. D. Bartlett}}},
  \bibinfo{author}{\bibnamefont{{M. W. Mitchell}}},
  \bibinfo{author}{\bibnamefont{{G. J. Pryde}}}, \bibnamefont{and}
  \bibinfo{author}{\bibnamefont{{H. M. Wiseman}}}, \bibinfo{journal}{Phys. Rev.
  A} \textbf{\bibinfo{volume}{80}}, \bibinfo{pages}{052114}
  (\bibinfo{year}{2009}).

\bibitem[{\citenamefont{{A. Sergeevich} et~al.}(2011)\citenamefont{{A.
  Sergeevich}, {A. Chandran}, {J. Combes}, {S. D. Bartlett}, and {H. M.
  Wiseman}}}]{sergeevich11}
\bibinfo{author}{\bibnamefont{{A. Sergeevich}}},
  \bibinfo{author}{\bibnamefont{{A. Chandran}}},
  \bibinfo{author}{\bibnamefont{{J. Combes}}},
  \bibinfo{author}{\bibnamefont{{S. D. Bartlett}}}, \bibnamefont{and}
  \bibinfo{author}{\bibnamefont{{H. M. Wiseman}}}, \bibinfo{journal}{Phys. Rev.
  A} \textbf{\bibinfo{volume}{84}}, \bibinfo{pages}{052315}
  (\bibinfo{year}{2011}).

\bibitem[{\citenamefont{{F. Husz\'ar} and {N. M. T. Houlsby}}(2012)}]{huszar12}
\bibinfo{author}{\bibnamefont{{F. Husz\'ar}}} \bibnamefont{and}
  \bibinfo{author}{\bibnamefont{{N. M. T. Houlsby}}}, \bibinfo{journal}{Phys.
  Rev. A} \textbf{\bibinfo{volume}{85}}, \bibinfo{pages}{052120}
  (\bibinfo{year}{2012}).

\bibitem[{\citenamefont{{C. E. Granade} et~al.}(2012)\citenamefont{{C. E.
  Granade}, {C. Ferrie}, {N. Wiebe}, and {D. G. Cory}}}]{granade12}
\bibinfo{author}{\bibnamefont{{C. E. Granade}}},
  \bibinfo{author}{\bibnamefont{{C. Ferrie}}},
  \bibinfo{author}{\bibnamefont{{N. Wiebe}}}, \bibnamefont{and}
  \bibinfo{author}{\bibnamefont{{D. G. Cory}}}, \bibinfo{journal}{New J. Phys.}
  \textbf{\bibinfo{volume}{14}}, \bibinfo{pages}{103013}
  (\bibinfo{year}{2012}).

\bibitem[{\citenamefont{{C. Ferrie} et~al.}(2013)\citenamefont{{C. Ferrie}, {C.
  E. Granade}, and {D. G. Cory}}}]{ferrie13}
\bibinfo{author}{\bibnamefont{{C. Ferrie}}}, \bibinfo{author}{\bibnamefont{{C.
  E. Granade}}}, \bibnamefont{and} \bibinfo{author}{\bibnamefont{{D. G.
  Cory}}}, \bibinfo{journal}{Quant. Inf. Proc.} \textbf{\bibinfo{volume}{12}},
  \bibinfo{pages}{611} (\bibinfo{year}{2013}).

\bibitem[{\citenamefont{{C. Ferrie}}(2014)}]{ferrie14a}
\bibinfo{author}{\bibnamefont{{C. Ferrie}}}, \bibinfo{journal}{Phys. Rev.
  Lett.} \textbf{\bibinfo{volume}{113}}, \bibinfo{pages}{190404}
  (\bibinfo{year}{2014}).

\bibitem[{\citenamefont{{N. Wiebe} et~al.}(2014{\natexlab{a}})\citenamefont{{N.
  Wiebe}, {C. E. Granade}, {C. Ferrie}, and {D. G. Cory}}}]{wiebe14a}
\bibinfo{author}{\bibnamefont{{N. Wiebe}}}, \bibinfo{author}{\bibnamefont{{C.
  E. Granade}}}, \bibinfo{author}{\bibnamefont{{C. Ferrie}}}, \bibnamefont{and}
  \bibinfo{author}{\bibnamefont{{D. G. Cory}}}, \bibinfo{journal}{Phys. Rev.
  Lett.} \textbf{\bibinfo{volume}{112}}, \bibinfo{pages}{190501}
  (\bibinfo{year}{2014}{\natexlab{a}}).

\bibitem[{\citenamefont{{N. Wiebe} et~al.}(2014{\natexlab{b}})\citenamefont{{N.
  Wiebe}, {C. E. Granade}, {C. Ferrie}, and {D. G. Cory}}}]{wiebe14b}
\bibinfo{author}{\bibnamefont{{N. Wiebe}}}, \bibinfo{author}{\bibnamefont{{C.
  E. Granade}}}, \bibinfo{author}{\bibnamefont{{C. Ferrie}}}, \bibnamefont{and}
  \bibinfo{author}{\bibnamefont{{D. G. Cory}}}, \bibinfo{journal}{Phys. Rev. A}
  \textbf{\bibinfo{volume}{89}}, \bibinfo{pages}{042314}
  (\bibinfo{year}{2014}{\natexlab{b}}).

\bibitem[{\citenamefont{{M. P. V. Stenberg} et~al.}(2014)\citenamefont{{M. P.
  V. Stenberg}, {Y. R. Sanders}, and {F. K. Wilhelm}}}]{stenberg14}
\bibinfo{author}{\bibnamefont{{M. P. V. Stenberg}}},
  \bibinfo{author}{\bibnamefont{{Y. R. Sanders}}}, \bibnamefont{and}
  \bibinfo{author}{\bibnamefont{{F. K. Wilhelm}}}, \bibinfo{journal}{Phys. Rev.
  Lett.} \textbf{\bibinfo{volume}{113}}, \bibinfo{pages}{210404}
  (\bibinfo{year}{2014}).

\bibitem[{\citenamefont{{C. E. Granade} et~al.}(2015)\citenamefont{{C. E.
  Granade}, {C. Ferrie}, and {D. G. Cory}}}]{granade15}
\bibinfo{author}{\bibnamefont{{C. E. Granade}}},
  \bibinfo{author}{\bibnamefont{{C. Ferrie}}}, \bibnamefont{and}
  \bibinfo{author}{\bibnamefont{{D. G. Cory}}}, \bibinfo{journal}{New J. Phys.}
  \textbf{\bibinfo{volume}{17}}, \bibinfo{pages}{013042}
  (\bibinfo{year}{2015}).

\bibitem[{\citenamefont{{M. P. V. Stenberg} et~al.}(2015)\citenamefont{{M. P.
  V. Stenberg}, {K. Pack}, and {F. K. Wilhelm}}}]{stenberg15}
\bibinfo{author}{\bibnamefont{{M. P. V. Stenberg}}},
  \bibinfo{author}{\bibnamefont{{K. Pack}}}, \bibnamefont{and}
  \bibinfo{author}{\bibnamefont{{F. K. Wilhelm}}}, \bibinfo{journal}{Phys. Rev.
  A} \textbf{\bibinfo{volume}{92}}, \bibinfo{pages}{063852}
  (\bibinfo{year}{2015}).

\bibitem[{\citenamefont{{M. P. V. Stenberg} et~al.}(2016)\citenamefont{{M. P.
  V. Stenberg}, {O. K\"ohn}, and {F. K. Wilhelm}}}]{stenberg16}
\bibinfo{author}{\bibnamefont{{M. P. V. Stenberg}}},
  \bibinfo{author}{\bibnamefont{{O. K\"ohn}}}, \bibnamefont{and}
  \bibinfo{author}{\bibnamefont{{F. K. Wilhelm}}}, \bibinfo{journal}{Phys. Rev.
  A} \textbf{\bibinfo{volume}{93}}, \bibinfo{pages}{012122}
  (\bibinfo{year}{2016}).

\bibitem[{\citenamefont{{Th. Hannemann} et~al.}(2002)\citenamefont{{Th.
  Hannemann}, {D. Reiss}, {Ch. Balzer}, {W. Neuhauser}, {P. E. Toschek}, and
  {Ch. Wunderlich}}}]{hannemann02}
\bibinfo{author}{\bibnamefont{{Th. Hannemann}}},
  \bibinfo{author}{\bibnamefont{{D. Reiss}}},
  \bibinfo{author}{\bibnamefont{{Ch. Balzer}}},
  \bibinfo{author}{\bibnamefont{{W. Neuhauser}}},
  \bibinfo{author}{\bibnamefont{{P. E. Toschek}}}, \bibnamefont{and}
  \bibinfo{author}{\bibnamefont{{Ch. Wunderlich}}}, \bibinfo{journal}{Phys.
  Rev. A} \textbf{\bibinfo{volume}{65}}, \bibinfo{pages}{050303(R)}
  (\bibinfo{year}{2002}).

\bibitem[{\citenamefont{{B. L. Higgins} et~al.}(2007)\citenamefont{{B. L.
  Higgins}, {D. W. Berry}, {S. D. Bartlett}, {H. M. Wiseman}, and {G. J.
  Pryde}}}]{higgins07}
\bibinfo{author}{\bibnamefont{{B. L. Higgins}}},
  \bibinfo{author}{\bibnamefont{{D. W. Berry}}},
  \bibinfo{author}{\bibnamefont{{S. D. Bartlett}}},
  \bibinfo{author}{\bibnamefont{{H. M. Wiseman}}}, \bibnamefont{and}
  \bibinfo{author}{\bibnamefont{{G. J. Pryde}}}, \bibinfo{journal}{Nature
  (London)} \textbf{\bibinfo{volume}{450}}, \bibinfo{pages}{393}
  (\bibinfo{year}{2007}).

\bibitem[{\citenamefont{{G. Y. Xiang} et~al.}(2011)\citenamefont{{G. Y. Xiang},
  {B. L. Higgins}, {D. W. Berry}, {H. M. Wiseman}, and {G. J.
  Pryde}}}]{xiang11}
\bibinfo{author}{\bibnamefont{{G. Y. Xiang}}},
  \bibinfo{author}{\bibnamefont{{B. L. Higgins}}},
  \bibinfo{author}{\bibnamefont{{D. W. Berry}}},
  \bibinfo{author}{\bibnamefont{{H. M. Wiseman}}}, \bibnamefont{and}
  \bibinfo{author}{\bibnamefont{{G. J. Pryde}}}, \bibinfo{journal}{Nat.
  Photon.} \textbf{\bibinfo{volume}{5}}, \bibinfo{pages}{43}
  (\bibinfo{year}{2011}).

\bibitem[{\citenamefont{{H. Yonezawa {\it et al.}}}(2012)}]{yonezawa12}
\bibinfo{author}{\bibnamefont{{H. Yonezawa {\it et al.}}}},
  \bibinfo{journal}{Science} \textbf{\bibinfo{volume}{337}},
  \bibinfo{pages}{1514} (\bibinfo{year}{2012}).

\bibitem[{\citenamefont{{K. S. Kravtsov} et~al.}(2013)\citenamefont{{K. S.
  Kravtsov}, {S. S. Straupe}, {I. V. Radchenko}, {N. M. T. Houlsby}, {F.
  Husz\'ar}, and {S. P. Kulik}}}]{kravtsov13}
\bibinfo{author}{\bibnamefont{{K. S. Kravtsov}}},
  \bibinfo{author}{\bibnamefont{{S. S. Straupe}}},
  \bibinfo{author}{\bibnamefont{{I. V. Radchenko}}},
  \bibinfo{author}{\bibnamefont{{N. M. T. Houlsby}}},
  \bibinfo{author}{\bibnamefont{{F. Husz\'ar}}}, \bibnamefont{and}
  \bibinfo{author}{\bibnamefont{{S. P. Kulik}}}, \bibinfo{journal}{Phys. Rev.
  A} \textbf{\bibinfo{volume}{87}}, \bibinfo{pages}{062122}
  (\bibinfo{year}{2013}).

\bibitem[{\citenamefont{{G. I. Struchalin} et~al.}(2016)\citenamefont{{G. I.
  Struchalin}, {I. A. Pogorelov}, {S. S. Straupe}, {K. S. Kravtsov}, {I. V.
  Radchenko}, and {S. P. Kulik}}}]{struchalin16}
\bibinfo{author}{\bibnamefont{{G. I. Struchalin}}},
  \bibinfo{author}{\bibnamefont{{I. A. Pogorelov}}},
  \bibinfo{author}{\bibnamefont{{S. S. Straupe}}},
  \bibinfo{author}{\bibnamefont{{K. S. Kravtsov}}},
  \bibinfo{author}{\bibnamefont{{I. V. Radchenko}}}, \bibnamefont{and}
  \bibinfo{author}{\bibnamefont{{S. P. Kulik}}}, \bibinfo{journal}{Phys. Rev.
  A} \textbf{\bibinfo{volume}{93}}, \bibinfo{pages}{012103}
  (\bibinfo{year}{2016}).

\bibitem[{\citenamefont{{M. West}}(1993)}]{west93}
\bibinfo{author}{\bibnamefont{{M. West}}}, \bibinfo{journal}{J. Roy. Stat. Soc.
  B Met.} \textbf{\bibinfo{volume}{55}}, \bibinfo{pages}{409}
  (\bibinfo{year}{1993}).

\bibitem[{\citenamefont{{N. J. Gordon} et~al.}(1993)\citenamefont{{N. J.
  Gordon}, {D. J. Salmond}, and {A. F. M. Smith}}}]{gordon93}
\bibinfo{author}{\bibnamefont{{N. J. Gordon}}},
  \bibinfo{author}{\bibnamefont{{D. J. Salmond}}}, \bibnamefont{and}
  \bibinfo{author}{\bibnamefont{{A. F. M. Smith}}}, \bibinfo{journal}{Radar and
  Signal Processing IEE Proc.-F} \textbf{\bibinfo{volume}{140}},
  \bibinfo{pages}{107} (\bibinfo{year}{1993}).

\bibitem[{J. Liu and M. West in, {\it Sequential Monte Carlo Methods in
  Practice}, edited by A. Doucet, N. Freitas, and N. Gordon (Springer, New
  York, 2001).({\natexlab{a}})}]{liu01}
J. Liu and M. West in, {\it Sequential Monte Carlo Methods in Practice}, edited
  by A. Doucet, N. Freitas, and N. Gordon (Springer, New York, 2001).

\bibitem[{\citenamefont{Lindblad}(1976)}]{lindblad76}
\bibinfo{author}{\bibfnamefont{J.}~\bibnamefont{Lindblad}},
  \bibinfo{journal}{Commun. Math. Phys.} \textbf{\bibinfo{volume}{48}},
  \bibinfo{pages}{119} (\bibinfo{year}{1976}).

\bibitem[{\citenamefont{{A. Blais} et~al.}(2004)\citenamefont{{A. Blais},
  {R.-S. Huang}, {A. Wallraff}, {S. M. Girvin}, and {R. J.
  Schoelkopf}}}]{blais04}
\bibinfo{author}{\bibnamefont{{A. Blais}}},
  \bibinfo{author}{\bibnamefont{{R.-S. Huang}}},
  \bibinfo{author}{\bibnamefont{{A. Wallraff}}},
  \bibinfo{author}{\bibnamefont{{S. M. Girvin}}}, \bibnamefont{and}
  \bibinfo{author}{\bibnamefont{{R. J. Schoelkopf}}}, \bibinfo{journal}{Phys.
  Rev. A} \textbf{\bibinfo{volume}{69}}, \bibinfo{pages}{062320}
  (\bibinfo{year}{2004}).

\bibitem[{\citenamefont{{M. Neeley} et~al.}(2008)\citenamefont{{M. Neeley}, {M.
  Ansmann}, {R. C. Bialczak}, {M. Hofheinz}, {N. Katz}, {E. Lucero},
  {A.O'Connell}, {H. Wang}, {A. N. Cleland}, and {J. M. Martinis}}}]{neeley08}
\bibinfo{author}{\bibnamefont{{M. Neeley}}}, \bibinfo{author}{\bibnamefont{{M.
  Ansmann}}}, \bibinfo{author}{\bibnamefont{{R. C. Bialczak}}},
  \bibinfo{author}{\bibnamefont{{M. Hofheinz}}},
  \bibinfo{author}{\bibnamefont{{N. Katz}}}, \bibinfo{author}{\bibnamefont{{E.
  Lucero}}}, \bibinfo{author}{\bibnamefont{{A.O'Connell}}},
  \bibinfo{author}{\bibnamefont{{H. Wang}}}, \bibinfo{author}{\bibnamefont{{A.
  N. Cleland}}}, \bibnamefont{and} \bibinfo{author}{\bibnamefont{{J. M.
  Martinis}}}, \bibinfo{journal}{Nat. Phys.} \textbf{\bibinfo{volume}{4}},
  \bibinfo{pages}{523} (\bibinfo{year}{2008}).

\bibitem[{\citenamefont{{M. Mariantoni {\it et
  al.}}}(2011{\natexlab{a}})}]{mariantoni11a}
\bibinfo{author}{\bibnamefont{{M. Mariantoni {\it et al.}}}},
  \bibinfo{journal}{Science} \textbf{\bibinfo{volume}{334}},
  \bibinfo{pages}{61} (\bibinfo{year}{2011}{\natexlab{a}}).

\bibitem[{\citenamefont{{M. Mariantoni {\it et
  al.}}}(2011{\natexlab{b}})}]{mariantoni11b}
\bibinfo{author}{\bibnamefont{{M. Mariantoni {\it et al.}}}},
  \bibinfo{journal}{Nat. Phys.} \textbf{\bibinfo{volume}{7}},
  \bibinfo{pages}{287} (\bibinfo{year}{2011}{\natexlab{b}}).

\bibitem[{J. Liu and M. West in, {\it Sequential Monte Carlo Methods in
  Practice}, edited by A. Doucet, N. Freitas, and N. Gordon (Springer, New
  York, 2001).({\natexlab{b}})}]{liu00}
J. Liu and M. West in, {\it Sequential Monte Carlo Methods in Practice}, edited
  by A. Doucet, N. Freitas, and N. Gordon (Springer, New York, 2001).

\bibitem[{D. Ascher, P. Dubois, K. Hinsen, J. Hugunin, and T. Oliphant,
  computer code \textsc{NUMERICAL PYTHON} 2001,
  \url{http://www.numpy.org}.()}]{ascher01}
D. Ascher, P. Dubois, K. Hinsen, J. Hugunin, and T. Oliphant, computer code
  \textsc{NUMERICAL PYTHON} 2001, \url{http://www.numpy.org}.

\bibitem[{E. Jones {\it et al.}, computer code \textsc{SCIPY}, 2001,
  \url{http://www.scipy.org}.()}]{jones01}
E. Jones {\it et al.}, computer code \textsc{SCIPY}, 2001,
  \url{http://www.scipy.org}.

\bibitem[{\citenamefont{{D. W. Berry} and {H. M. Wiseman}}(2000)}]{berry00}
\bibinfo{author}{\bibnamefont{{D. W. Berry}}} \bibnamefont{and}
  \bibinfo{author}{\bibnamefont{{H. M. Wiseman}}}, \bibinfo{journal}{Phys. Rev.
  Lett.} \textbf{\bibinfo{volume}{85}}, \bibinfo{pages}{5098}
  (\bibinfo{year}{2000}).

\bibitem[{\citenamefont{{D. G. Fischer} and {M. Freyberger}}(2000)}]{fischer00}
\bibinfo{author}{\bibnamefont{{D. G. Fischer}}} \bibnamefont{and}
  \bibinfo{author}{\bibnamefont{{M. Freyberger}}}, \bibinfo{journal}{Phys.
  Lett. A} \textbf{\bibinfo{volume}{273}}, \bibinfo{pages}{293}
  (\bibinfo{year}{2000}).

\bibitem[{\citenamefont{{D. W. Berry} et~al.}(2001)\citenamefont{{D. W. Berry},
  {H. M. Wiseman}, and {J. K. Breslin}}}]{berry01}
\bibinfo{author}{\bibnamefont{{D. W. Berry}}},
  \bibinfo{author}{\bibnamefont{{H. M. Wiseman}}}, \bibnamefont{and}
  \bibinfo{author}{\bibnamefont{{J. K. Breslin}}}, \bibinfo{journal}{Phys. Rev.
  A} \textbf{\bibinfo{volume}{63}}, \bibinfo{pages}{053804}
  (\bibinfo{year}{2001}).

\bibitem[{M. A. Nielsen and I. L. Chuang, {\it Quantum Computation and Quantum
  Information}, (Cambridge University Press, Cambridge, England,
  2000).()}]{nielsen00}
M. A. Nielsen and I. L. Chuang, {\it Quantum Computation and Quantum
  Information}, (Cambridge University Press, Cambridge, England, 2000).

\end{thebibliography}
\end{document}